\documentstyle[12pt]{article}
\def\be{\begin{equation}}
\def\ee{\end{equation}}
\def\beqn{\begin{eqnarray}}
\def\eeqn{\end{eqnarray}}

\begin{document}
\begin{titlepage}
\title{Orthogonality relations and supercharacter formulas
of $U(m|n)$ representations} \author{ Jorge Alfaro \thanks{e-mail
address:
jalfaro@lascar.puc.cl .}
\\ Facultad de F\'\i sica\\Universidad Cat\'olica de Chile\\Casilla 306,
Santiago 22, Chile\\  {} \\ Ricardo Medina \thanks{e-mail address:
rmedina@power.ift.unesp.br} \\ Instituto de F\'\i sica Te\'orica\\
Universidade
Estadual Paulista\\ Rua Pamplona 145\\ 01405-900 S\~{a}o
Paulo, Brasil\\
{} \\
and \\ {} \\ Luis F.  Urrutia \\ Instituto de Ciencias Nucleares\\
Universidad Nacional
Aut\'onoma de M\'exico\\ Circuito Exterior, C.U.\\ Apartado Postal
70-543,
04510
M\'exico, D.F.  } 
\maketitle

\newpage

\begin{abstract}

In this paper we obtain the orthogonality relations for the supergroup
$U(m|n)$,
which are remarkably different from the ones for the $U(N)$ case.
We extend our
results for ordinary representations, obtained some time ago, to
the case of
complex conjugated and mixed representations. Our results are
expressed in terms
of the Young tableaux notation for irreducible representations. We
use the
supersymmetric Harish-Chandra-Itzykson-Zuber integral and the character
expansion technique as mathematical tools for deriving these
relations. As a
byproduct  we also  obtain closed expressions for the supercharacters
and dimensions  of  some particular irreducible $U(m|n)$ representations.
A new way of labeling  the $U(m|n)$ irreducible representations in
terms of
$m + n$ numbers is proposed. Finally, as a corollary of our results,
new identities among the dimensions
of the irreducible representations of the  unitary group $U(N)$
 are presented.

\end{abstract}
\vfill 
\begin{flushleft} IFT-P.014/97
\end{flushleft}
\end{titlepage}

\newpage
\baselineskip=20 pt
\section{Introduction}

In recent times there has been an enormous amount of work devoted to the
understanding of random surfaces and statistical systems on random
surfaces.
The range of application of these ideas include non-critical string
theory as
well as Quantum Chromodynamics (QCD) in the large $N$ limit.
Progress in this
area has been possible because the mathematical knowledge on random
matrices
has
increased dramatically in the last fifteen years \cite{mehta1}.

An important mathematical  concept  that appears naturally in the
discussion of
random matrices is  the  integration over the unitary group, which basis
are well understood in the literature.  A distinguished particular
case of such
integrals,
the Harish-Chandra-Itzykson-Zuber (HCIZ)  integral  \cite{HC, IZ},
has been
applied to the solution of
the Two matrix
model \cite{IZ, mehta2} and, more recently, to the
Migdal-Kazakov model
of "induced QCD" \cite{migkaz}.  In a different context, it has also been
applied to the study of phase transitions in nematic liquids
\cite{Mulder} .
The HCIZ integral   can also be considered a powerful  alternative
 tool for
deriving results regarding
the representation theory of the group $U(N)$.

On the other hand, since its discovery,  there has been  considerable
 expectation that
supersymmetry might
play an important role in the physical world . This hope  has
motivated, on one
hand,
the extension of many important physical ideas to the
supersymmetric world
\cite{susy}. A related example
of direct interest to us is the case of random supermatrices and
supermatrix
models
\cite{gilbert}. On the other hand, this expectation  has  also
contributed  to
the study and development of  the associated mathematical tools:
supermanifolds,
differential and integral calculus over a Grassmann algebra, differential
geometry over a supermanifold, superalgebras and Lie supergroups
among others
\cite{berezin1,Dewitt}.

This paper
deals with  the integration properties of the  unitary supergroup
$U(m|n)$.
For our purposes
we will  work in  a representation of   this  supergroup
given by  the set of all  $ (m+n) \times (m+n)$  supermatrices
$U=[ U_{AB}]$,
such that
$U U^\dagger= 1$,  endowed with the operation of supermatrix
multiplication.

The issue of defining an invariant
integral for supergroups has been discussed previously in
\cite{berezin1},
\cite{Cornwell} and \cite{ Fronsdal}, among other references.  In
Ref.\cite{Cornwell}  the problem
is solved  by defining the invariant integral over a Lie supergroup
as equal to
that over its related
Lie group and subsequently using the  theory of invariant (Haar)
integrals for
topological groups \cite {Nachbin, Hewitt}.  References
\cite{berezin1}
and \cite{Fronsdal} are
on the line of a  physicist approach, by keeping the grassmannian
character  of
the integration
volume element.  We adhere to the  last point of view  and we
introduce an
integration  measure $[dU]$  based on the Berezin  integration
properties of
the independent elements of the supermatrix. As first noted by Berezin
\cite{berezin1} and also in the paper by Yost in  Ref.
\cite{gilbert}, the
unusual  property
$\int  [dU]=0 $ will hold, thus making the calculation of the
orthogonality
relations a much  more  involved issue. Orthogonality relations for
$U(m|n)$
have been
previously obtained by Berezin \cite{berezin1} and formulated in
terms of   the
classification of the
representations of the supergroup via the Cartan approach .
Instead, we use
the Young
tableaux  method for classifying the  irreducible representations
for $U(m|n)$
\cite{BARS0}. We have not studied  the relation between  Berezin's
result and
our way  of presenting the orthogonality relations, which  are
derived  using
a completely different approach.
Our method of calculation is based on the result obtained  for the
supersymmetric extension
of the HCIZ integral \cite{Nosotros1,GUHR2}, together with the use
of character
expansion techniques.

The paper is organized as follows :  Sections \ref{INTSM} and   
\ref{BPSR}  are
basically a  brief
review of
supermatrices and supergroup representations, respectively,
designed to make
the presentation self-contained and also to  introduce our notation and
conventions.

Sections \ref{ORFU} and \ref{ACM}  deal with the orthogonality relations
for the irreducible representations of $U(m|n)$.  Besides  the  expected
product of
Kronecker deltas, these relations include a  representation dependent
coefficient
$\alpha_{\{t\}}$ which  calculation, for all three  types of
representations of
the supergroup (ordinary, complex conjugated  and mixed), is the
main subject
of this section. We  show that
in the case of the mixed representations  these
coefficients
 can be  written  in terms of those corresponding to the ordinary
representations. Some examples are presented in the Appendix \ref{Alfas}
The unusual
property of the integration measure mentioned  above  has also the
consequence
that this coefficient is non-zero only for a class  of
representations  which
are completely  identified in our approach. Some preliminary
results regarding
this issue
were previously presented in Ref.\cite{Nosotros2}. Here we have
completed the
determination of
the coefficients
 $\alpha_{\{t\}}$ for the cases that were  missing  in
\cite{Nosotros2} and we
also  give  a  more detailed version of our calculation. Closed
formulas for
the dimensions and supercharacters of ordinary and complex conjugate
 representations with
$\alpha_{\{t\}}
\neq 0$ are also presented.

The restriction  $U(m|n) \rightarrow U(m)$  correctly reproduces
the result
$ \alpha_{\{t\}} \rightarrow {1 \over  d_{\{t\}}} $ in the orthogonality
relations, where $ d_{\{t\}}$ is the dimension of the corresponding
representation.  In this way, our expressions for the $\alpha_{\{t\}}$
coefficients
of  the mixed representation in terms of those of the ordinary
representations, turn into identities for the corresponding
dimensions of the
$U(m)$ representations. Up to our knowledge, these
identities were not known before and they are presented in Section
\ref{IDU}. Some
specific examples
can be read off  in  the Appendix  \ref{Alfas},  after the  
replacement $\alpha
\rightarrow {1\over d}$ is made.

Section \ref{RUR} contains a proposal  to label
the irreducible representations of $U(m|n)$,  in terms of a finite
array of
$m+n$ numbers, not all necessarily independent, instead of giving an
arbitrary
large array of numbers ( that can contain infinite numbers in principle)
corresponding to the number of  boxes
in the rows  of the associated  Young tableau. The possible
advantages of this relabelling are not further explored.

Appendix  \ref{SUSYHCIZ} contains a brief review of the  
supersymmetric HCIZ
integral
which
 result is the basic tool  used in  our calculations. The remaining
Appendices
are the detailed calculations of some expressions in the main text,
together
with
the statement of useful relations  which are also used along the paper.

Finally, Tables I (II) in the Appendix  \ref{charactertable}    
contain  a list
of
characters
and dimensions of representations of the group $GL(N)$ (
supercharacters and
dimensions of representations of the supergroup $GL(m|n)$ ) which are an
extended version of those found in Ref. \cite{IZ}.

\section{ Introduction to supermatrices}
\label{INTSM}

Supergroups can be conveniently represented
by matrices
acting on a superspace (supermatrices).  To this end we briefly
review some of
the basic properties of the linear algebra defined over a Grassmann
algebra.
This sets the stage for the rest of the paper and also fixes our
notation.  For
a more detailed and complete discussion on these matters the reader
is referred
to Refs.\cite { berezin1, Dewitt}.

Let us consider a superspace with coordinates $z^P=(q^i, \theta^\alpha),
i=1,\dots m, \break \alpha=1, \dots, n$ such that the $q^i$'s
($\theta^\alpha$'s) are
even \ (odd) elements of a Grassmann algebra.  This means that $z^P z^Q=
(-1)^{\epsilon(P) \epsilon(Q)} z^Q z^P$, where $\epsilon(P)$ is the
Grassmann
parity of the index $P$ defined by $\epsilon(i)=0,\ mod(2);
\epsilon(\alpha)=
1,\ mod(2)$.The above multiplication rule implies in
particular that any
odd element of the Grassmann algebra has zero square, i.e.  it is
nilpotent. Also we have that $\epsilon(z^{P_1}z^{P_2} \dots
z^{P_k})= \sum
\epsilon(P_i)$.

Supermatrices are arrays that act linearly on the supercoordinates
leaving
invariant the partition among even and odd coordinates.  To be more
specific,
the supercoordinates can be thought as forming an $(m+n)\times 1$
column vector
with the first $m$ entries\ (last $ n$ entries) being even  (odd)
elements of
the Grassmann algebra.  In this way, an $(m+n)\times (m+n)$
supermatrix is an
array written in the partitioned block form
\begin{eqnarray}
M= \left( \begin{array}{ll} A_{m \times m} & B_{m \times n} \\
                            C_{n \times m} & D_{n \times n}
          \end{array} \right),
\label{block-form}
\end{eqnarray}
where the constituent matrices have components $A_{ij},\
B_{i\alpha},\ C_{\alpha i}$ and $D_{\alpha\beta}$.  Besides, $A_{ij},\
D_{\alpha\beta} \ (B_{i\alpha}, \ C_{\alpha i})$ are even \ (odd)
elements of
the Grassmann algebra in such a way that the parity array of the
supercoordinate
vector columnn is preserved under supermatrix multiplication of
that vector.
The parity of any supermatrix element is
$\epsilon(M_{PQ})= \epsilon(P)+\epsilon(Q)$. The addition and 
multiplication of supermatrices according to the rules
\begin{eqnarray*}
(M_1+M_2)_{PQ}= (M_1)_{PQ}+(M_2)_{PQ},\ \
(M_1M_2)_{PQ}= \sum_R (M_1)_{PR}(M_2)_{RQ},
\end{eqnarray*}
is such that it
produces again a supermatrix.  The inverse of a supermatrix can be
constructed
in block form, in complete analogy with the classical case and it is well
defined provided $A^{-1}$ and $D^{-1}$ exist.  The inverse of these
even matrices is
calculated in the standard way.

The basic invariant of a supermatrix under similarity
transformations is the
supertrace \begin{eqnarray*} Str(M)=Tr(A)-Tr(D)=\sum_{P=1}^{m+n}
(-1)^{\epsilon(P)}M_{PP},\label{str} \end{eqnarray*} which is
defined so that
the cyclic property $Str(M_1M_2)=Str(M_2M_1)$ is fullfilled for arbitrary
supermatrices $M_1, M_2$.  The above definition of the supertrace
leads to the
construction of the superdeterminant in the form $Sdet(M)=exp[
Str(ln M)]$,
which is explicitly given by the following two equivalent forms
\cite{alemanes}
\begin{eqnarray} Sdet(M)= \frac {det(A-B D^{-1} C)}{ det(D)}= \frac
{det(A)}{det(D-C A^{-1} B)}.  \end{eqnarray} The above expression
is written
only in terms of determinants of even matrices in such a way that
the determinant
has its usual
meaning.  The superdeterminat has the multiplicative property
$Sdet(M_1M_2)=
Sdet(M_1) Sdet(M_2)$.

The definition of the adjoint supermatrix follows the usual steps
by requiring
the identity $(y^{P*} M_{PQ} z^Q)^*=z^{P*}{M^\dagger}_{PQ} y^Q$, for an
arbitrary bilinear form in the complex supercoordinates $y^P$, where $*$
denotes
complex conjugation.  Since the usual definition of complex
conjugation in a
Grassmann algebra, $(y^P y^Q)^*=y^{Q*} y^{P*}$, reverses the order of
the factors
without introduccing any sign factor , we have the result
${M^\dagger}_{PQ}=
{M_{QP}}^*$ as in the standard case.

A hermitian $(m+n)\times(m+n)$ supermatrix $M$ is such that
$M^\dagger=M$ and
it
has $(m+n)^2$ real independent components.  The following
properties are also
fullfilled:\ (i)\ $(M^\dagger)^\dagger=M$, \ (ii) \ $(M_1M_2)^\dagger=
{M_2}^\dagger {M_1}^\dagger$ and (iii) \ $Sdet(M^\dagger)= {Sdet(M)}^*$.

A unitary $(m+n)\times(m+n)$ supermatrix $U$ is such that $UU^\dagger=
U^\dagger
U=I$ ( where $I$ is the identity supermatrix) and also has $(m+n)^2$ real
independent components, which have the additional property that
$(SdetU)(SdetU)^*=1$.  The set of all $(m+n)\times(m+n)$ unitary
supermatrices, under the operation of
supermatrix
multiplication,
constitutes  the supergroup $U(m|n)$.
Under very general conditions \cite{JAP},  hermitian supermatrices can be
diagonalized by superunitary transformations, thus introducing the
corresponding eigenvalues.  Our notation is such that the first $m$
eigenvalues
of an $ (m+n) \times (m+n)$ hermitian supermatrix $M$ are denoted by
$\lambda_i$,
while the remaining $n$ eigenvalues are denoted by
${\bar\lambda}_{\alpha}$.
Such partition is characterized by the following parity assignment of the
corresponding eigenvector components $V_P,\bar V_P:  \epsilon(V_P)=
\epsilon(P),\epsilon(\bar
V_P)=\epsilon(P)+1$, which are called eigenvectors of the first and
second
class
respectively. Thus, a diagonalizable hermitian supermatrix can be
decomposed as $M = U \Lambda U^\dagger$, where $U$ is a unitary
supermatrix
(which is built from the eigenvectors of $M$) and
\begin{eqnarray}
\Lambda = \left ( \begin{array}{cc}
                  \lambda_{m \times m} & 0 \\
                  0                    & \bar{\lambda}_{n \times n}
                  \end{array}
          \right ),
\label{diagonalsup}
\end{eqnarray}
is a diagonal supermatrix, $\lambda_{m \times m}$
($\bar{\lambda}_{n \times n}$) being an $m \times m$ ($n \times n$)
diagonal
matrix with components $\lambda_i$ ($\bar{\lambda}_\alpha$).

\section{Basic Properties of supergroup representations}
\label{BPSR}







Supergroups will be represented by linear operators ${\tilde D}(g)$
acting on
some vector space with basis $ \{ \Phi_I \}$.  Linearity is defined
by
${\tilde D}(g) \left(\Phi_I \alpha +\Phi_J \beta \right) = \left( {\tilde
D}(g)\Phi_I\right) \alpha + \left( {\tilde D}(g)\Phi_J \right)
\beta$, where
$\alpha$ and $\beta$ are arbitrary Grassmann numbers.  An
alternative choice is
produced by having the factors to the left.

The action
\begin{equation}
{\tilde D}(g) \left(\Phi_I \right) = \sum_{J}
\Phi_J {\cal D}^{(t)}_{JI}(g),
\label{REP}
\end{equation}
defines a
representation $(t)$ of the supergroup.

In spite of the use of Grassmann variables in our definition of
linearity,
the representation property
${\cal D}^{(t)}_{JI}(g_1*g_2)=\sum_K {\cal D}^{(t)}_{JK}(g_1) {\cal
D}^{(t)}_{KI}(g_2)$ is verified, thus showing that the
representation of the
supergroup elements in
terms of supermatrices respects the multiplication rule of supermatrices.

The above linearity convention applies also to the action of group
operators
acting upon vectors of the space.  Let us consider the vector
$\Psi=\sum_K
\Phi_K \alpha_K$ with components $\alpha_K$.  Then we have
\begin{equation}
{\tilde D}(g)\left(\Psi \right)=\sum_K {\tilde D}(g)\left(\Phi_K
\right)\alpha_K=\sum_{K, L} \Phi_L {\cal D}^{(t)}_{LK}(g)\alpha_K,
\end{equation}
which is consistent with the representation of a vector as a column with
entries
$\alpha_K$, together with the representation of the action of a
group element
upon such vector as the multiplication of the corresponding
supermatrix by the
respective columm.

Now, let us recall that there are two fundamental representations of
$U(m|n)$ : The ordinary one (or undotted),
${\cal D} _{ij}^{\begin{picture}(6,6)(0,0)
\put(0,0){\framebox(6,6){$\space$}}
\end{picture}} (U) = U_{ij}$, and the complex conjugate one (or dotted),
${\cal  D}_{ij}^{\begin{picture}(6,6)(0,0)
\put(0,0){\framebox(6,6){$\cdot$}} \end{picture}} (U) = \bar{U}_{ij} =
(-1)^{\epsilon_i (\epsilon_i + \epsilon_j)} U_{ij}^*$ \cite{BARS0}.
It is a direct calculation to show that
$\bar{U}$ is a unitary supermatrix and also that
$\bar{(UV)}={\bar{U}}{\bar{V}}$
for arbitrary $U(m|n)$ supermatrices, thus showing that the bar operation
constitutes indeed a representation of the supergroup.

Using the fundamental representations,
 three types of
irreducible representations $\{t\}$ are built : ordinary (undotted)  
$\{u\}$,
complex conjugated
(dotted ) $\{{\dot v}\}$
and mixed
$\{{\dot v}\}|\{u\}$, which we do, in analogy to the $U(N)$ case, according
 to the conventions in Ref.
\cite{BARS1}. In particular we
have
$\{u\}$=$\{{\dot 0}\}|\{u\}$ and $\{{\dot v}\}= \{{\dot v}\}|\{0\}$.

Contrary to what happens in the $SU(N)$ case, the dotted and undotted
 representations cannot be related
through
an epsilon symbol \cite{BARS1}, so they are not equivalent.

We will label the irreducible representations by means of the Young  
tableaux
notation. Thus, an undotted irreducible representation $\{t\}$ will
be characterized by the non negative integers $(t_1, t_2, \ldots ,t_k)$,
where $t_1 \geq t_2 \geq \ldots \geq t_k$ are the number of
boxes in the corresponding rows  of the tableau. For the moment we assume
that there is no restriction upon the number of $t_i$'s characterizing
the tableau.
Pictorically the tableau will
look like

\begin{center}
\begin{picture}(72,60)(0,0)
\put(0,0){\framebox(12,12){$\space$}}
\put(0,12){\framebox(12,24){$\vdots$}}
\put(12,12){\framebox(24,24){$\vdots$}}
\put(0,36){\framebox(12,12){$\space$}}
\put(12,36){\framebox(12,12){$\space$}}
\put(24,36){\framebox(12,12){$\space$}}
\put(36,36){\framebox(12,12){$\space$} $t_2$}
\put(0,48){\framebox(12,12){$\space$}}
\put(12,48){\framebox(12,12){$\space$}}
\put(24,48){\framebox(12,12){$\space$}}
\put(36,48){\framebox(12,12){$\space$}}
\put(48,48){\framebox(12,12){$\space$}}
\put(60,48){\framebox(12,12){$\space$} $t_1$}
\put(12,0){\framebox(12,12){$\space$}  $t_k.$}
\end{picture}
\end{center}
\begin{eqnarray}
\label{tableau1}
\end{eqnarray}
The supermatrix representation ${\cal D}^{\{t\}}(g)$ will then be an
$(m_{\{t\}} +n_{\{t\}})\times(m_{\{t\}}+n_{\{t\}})$ supermatrix
written in the standard form (\ref{block-form}),
consisting of elements
${\cal D}^{\{t\}}_{JI}(g)$.

So besides the undotted representations pictorically shown in  
(\ref{tableau1})
the dotted and mixed ones will look like

\begin{center}
\begin{picture}(48,48)(0,0)
\put(-36,15){$\{\dot{v}\}=$}
\put(0,0){\framebox(12,12){$\cdot$}}
\put(0,12){\framebox(12,12){$\cdot$}}
\put(0,24){\framebox(12,12){$\cdot$}}
\put(12,24){\framebox(12,12){$\cdot$}}
\end{picture}
\end{center}
\begin{eqnarray}
\label{v-tableau}
\end{eqnarray}
and
\begin{center}
\begin{picture}(60,48)(0,0)
\put(-88,21){$\{\dot{v}\}|\{u\}=$}
\put(60,21){$.$}
\put(0,0){\framebox(12,12){$\space$}}
\put(-12,12){\framebox(12,12){$\cdot$}}
\put(0,12){\framebox(12,12){$\space$}}
\put(-12,24){\framebox(12,12){$\cdot$}}
\put(0,24){\framebox(12,12){$\space$}}
\put(12,24){\framebox(12,12){$\space$}}
\put(24,24){\framebox(12,12){$\space$}}
\put(-24,36){\framebox(12,12){$\cdot$}}
\put(-12,36){\framebox(12,12){$\cdot$}}
\put(0,36){\framebox(12,12){$\space$}}
\put(12,36){\framebox(12,12){$\space$}}
\put(24,36){\framebox(12,12){$\space$}}
\put(36,36){\framebox(12,12){$\space$}}
\end{picture}
\end{center}
\begin{eqnarray}
\label{mixed-tableau}
\end{eqnarray}
Now, we observe
that in the
case of ordinary groups the determinant $det(U)$ provides a one
dimensional
representation which can be constructed as a completely antisymmetrized
product of
fundamental representations.  In the case of supergroups, the
superdeterminant
$Sdet(U)$ provides also a one dimensional representation which,
nevertheless,
cannot be constructed in terms of the fundamental representations.
This is
because the superdeterminant is a non polynomial function of the
eigenvalues.

When considering tensor products of the fundamental representations
we define
\begin{eqnarray}
\underbrace{({\cal D}^{\begin{picture}(6,6)(0,0)
\put(0,0){\framebox(6,6){$\space$}}
\end{picture}} \otimes \ldots  \otimes
{\cal D}^{
\begin{picture}(6,6)(0,0)
\put(0,0){\framebox(6,6){$\space$}}
\end{picture}})}_{p \ times} (U) = \oplus_{\{t\},|t|=p} \
\sigma_{\{t\}} {\cal D}^{\{t\}} (U),
\label{SIGMA}
\end{eqnarray}
where $|t|$ represents the number of boxes of the representation
$\{t\}$ and
$\sigma_{\{t\}}$ is a
Clebsch-Gordan coefficient which represents the number of times that the
irreducible representation ${\{t\}}$ is contained in the above
tensor product. It may be calculated using the Young tableaux rules for
the tensor product of representations in (\ref{SIGMA}) or
alternatively using
the formula (see Chapter 7, formula $(5.21)$ of Ref. \cite{Barut})
\begin{eqnarray}
\sigma_{\{t\}} = |t|! \frac{\Delta(t_1+k-1, t_2+k-2, \ldots, t_k)}
{\prod_{p=1}^{k} (t_{p}+k-p)!},
\label{sigmanueva}
\end{eqnarray}
in terms of the tableau labels given in (\ref{tableau1}), where
\begin{eqnarray}
\Delta(l_1, \ldots, l_k) = \prod_{i>j=1}^k (l_i - l_j)
\end{eqnarray}
is the Vandermonde determinant.

Some values of $\sigma_{\{t\}}$ are given in the tables in
Appendix \ref{charactertable}.
In particular, Eq.(\ref{SIGMA}) implies that
\begin{equation} (str U)^p =
\sum_{\{t\},|t|=p} \sigma_{\{t\}}{s \chi}_{\{t\}}(U),
\label{formula}
\end{equation}
where the
supercharacter of representation $\{t\}$ is
\begin{eqnarray}
s \chi_{\{t\}}(U) = str({\cal D}^{\{t\}}(U)),
\label{SUPERCHARACTER}
\end{eqnarray}
and where we have also used the property that the supercharacter of
tensor product of representations equals the product of the corresponding
supercharacters. An explicit formula for $s \chi_{\{t\}}(U)$ in terms
 of supermatrix $U$ is given in Appendix \ref{charactertable}.

Let us emphasize that  the  Grassmannian character of the
supermatrices involved  introduces further sign factors with
respect to the
classical case
in the case of tensor products.
Let us illustrate this point with
the direct
product of two fundamental undotted representations.  The
corresponding basis
vectors are $\Psi_i \Phi_j$ which are rotated to $\Psi'_k\Phi'_l$ by the
independent actions of the supergroup $\Psi'_k=\Psi_i U_{ik}$ and
$\Phi'_l=\Phi_j
U_{jl}$.  Looking for the transformation of the product we have
\begin{equation}
\Psi'_k\Phi'_l=\left(\Psi_i U_{ik}\right)\left(\Phi_j U_{jl}\right)
=\Psi_i\Phi_j
\left((-1)^{\epsilon_j(\epsilon_k+\epsilon_i)}U_{ik}U_{jl}\right),
\end{equation}
which identifies \begin{equation} \left({\cal
D}^{\begin{picture}(6,6)(0,0) \put(0,0){\framebox(6,6){$\space$}}
\end{picture}}
\times {\cal D}^{\begin{picture}(6,6)(0,0)
\put(0,0){\framebox(6,6){$\space$}}
\end{picture}} \right)_{ij,kl}(U)=
(-1)^{\epsilon_j(\epsilon_k+\epsilon_i)}U_{ik}U_{jl}.\label{prod}
\end{equation}

It is a direct calculation to verify that this assigment
constitutes indeed a
representation of the supergroup.

The expression (\ref{prod}) can be generalized for an arbitrary
tensor product
\begin{eqnarray}
\underbrace{({\cal D}^{\begin{picture}(6,6)(0,0)
\put(0,0){\framebox(6,6){$\space$}}
\end{picture}} \times \ldots  \times
{\cal D}^{
\begin{picture}(6,6)(0,0)
\put(0,0){\framebox(6,6){$\space$}}
\end{picture}})_{IJ}}_{p \ times} (U)= (-1)^{\epsilon_{j_1}(
\epsilon_{i_2} +
\epsilon_{j_2} +
\dots + \epsilon_{i_p} + \epsilon_{j_p}) + \dots} & \nonumber \\
\times (-1)^{ \epsilon_{j_{p-1}} ( \epsilon_{i_p} +
\epsilon_{j_p})} U_{i_1
j_1}  U_{i_2 j_2}
\dots U_{i_p j_p},   &
\label{tensprod}
\end{eqnarray}
with $I= \{ i_1, i_2, \dots, i_p\}, J= \{ j_1, j_2, \dots, j_p\}$.

As we mentioned before, the construction of the irreducible tensor
representations symmetrized
according to a specific Young tableau, to which we referred in
(\ref{tableau1}), (\ref{v-tableau}) and (\ref{mixed-tableau}),
 proceeds in complete analogy to the
$U(N)$ case,  as stated in Ref. \cite{BARS2}.
In particular, the corresponding supercharacters are exactly those
of $U(N)$
with the trace
replaced by a supertrace (see appendix \ref{charactertable} for  
some examples).

\section{Orthogonality relations for $U(m|n)$}
\label{ORFU}

\subsection{Unitary supergroup measure}
\label{ORFU1}

For finding the orthogonality relations we will make use of the
Schur's lemma,
extended to the case of continuous supergroups. We will have to deal with
supergroup integration and for this reason we briefly refer to the
unitary
supergroup measure.

In general, the supergroup measure must be left and
right-invariant under the supergroup action. In the case of
$U(m|n)$ it is
defined by
\begin{equation}
[dU] = \mu \prod_{P,Q=1}^{m+n} dU_{PQ} dU_{PQ}^* \ \delta
(UU^\dagger - I),
\end{equation}
 where the $\delta$-function really means the product of
$(m+n)^2$ unidimensional $\delta$-functions corresponding to the
independent
constraints set by the condition $UU^\dagger = I$. The integration
over each
Grassmann
valued element $dU_{PQ}$ is defined according to the standard
Berezin's rules.
 The  arbitrary non null constant  $\mu$ will be  fixed  from  the
convention
adopted  for our
normalization of the supersymmetric HCIZ  integral.  It is
important to observe that although the above measure contains odd
differentials
and odd variables, it has $0$ Grasmann parity and therefore behaves as
an even Grassmann variable (commutes with everything).

\subsection{General Form of the Orthogonality relations}
\label{ORFU2}

In order to derive the general form of the orthogonality relations
 we apply Schur's lemma to the
quantity
${\cal X}^{\{s\},\{t\}}_{IL}=\int [dU]{\cal D}^{\{s\}}_{IJ}(U)
X_{JK}{\cal
D}^{\{t\}}_{KL}(U^{-1})$, where
$X_{JK}$ is an arbitrary supermatrix. We are assuming sum over repeated
indices.
Multiplying this expression to the left by the arbitrary element ${\cal
D}^{\{s\}}_{RI}(S)$ and using the composition property of the
representation we
obtain
\begin{equation}
{\cal D}^{\{s\}}_{RI}(S)
{\cal  X}^{\{s\},\{t\}}_{IL}=
\int [dU] {\cal D}^{\{s\}}_{RJ}(SU) X_{JK} {\cal D}^{\{t\}}_{KL}(U^{-1}).
\end{equation}
Here we used the fact that $[dU]$ behaves like an even Grassmann
variable.
Next we rewrite ${\cal D}^{\{t\}}_{KL}(U^{-1})= {\cal
D}^{\{t\}}_{KM}((SU)^{-1}){\cal D}^{\{t\}}_{ML}(S)$ and substitute this
expression in the previous equation, obtaining \begin{equation} {\cal
D}^{\{s\}}_{RI}(S){\cal X}^{\{s\},\{t\}}_{IL}= \left(\int [dU] {\cal
D}^{\{s\})}_{RJ}(SU) X_{JK}{\cal D}^{\{t\}}_{KM}((SU)^{-1})\right) {\cal
D}^{\{t\}}_{ML}(S).
\end{equation}
{}From  the invariance of the
measure under
left multiplications we realize that the quantity in brackets is
precisely
${\cal X}^{\{s\},\{t\}}_{RM}$ and therefore we obtain that
${\cal D}^{\{s\}}(S){\cal X}^{\{s\},\{t\}} =
{\cal X}^{\{s\},\{t\}}{\cal D}^{\{t\}}(S)$. Then, in analogy with
the ordinary
case, we have that : (i)  if $\{s\} \neq \{t\}$
then ${\cal X}^{\{s\},\{t\}} = 0$, and  (ii) if $\{s\} = \{t\}$ then
${\cal X}^{\{s\},\{s\}}$ is a multiple of the $sd_{\{s\}}$-dimensional
identity supermatrix (where $sd_{\{s\}}$ is the dimension of the $\{s\}$
 representation). Thus,
\begin{eqnarray}
{\cal X}^{\{s\},\{t\}}_{IL} (X) =
\int[dU] {\cal D}^{\{s\}}_{IJ}(U) X_{JK} {\cal D}^{\{t\}}_{KL}(U^{-1})=
\alpha^{\{s\}}(X)\delta^{\{s\},\{t\}} \delta_{IL}^{\{s\}},
\label{sidentity}
\end{eqnarray}
where the coefficient $\alpha$
depends upon
the arbitrary supermatrix $X$.  We can prove that the above equation is
invariant under the rotation $X'={\cal D}^{\{s\}}X{{{\cal
D}}^{\{s\}}}^{-1}$,
for a given
representation $\{s\}$,  in virtue of the composition properties of a
representation together with the invariance of the measure with
respect to right
multiplication.  This means that $\alpha^{\{s\}}(X)$ must be an
invariant under
similarity transformations,  which is linear in $X$.  The only
possibility is
that
$\alpha^{\{s\}}(X) =\alpha^{\{s\}} str X $, where $\alpha^{\{s\}}$
is now a
numerical coefficient.  So, in (\ref{sidentity}) we have obtained
an equality
between
two linear expressions of the $X_{IK}$'s.
Taking care of the Grasmannian
character of the
indices involved, the comparison  of the coefficients of the fully
independent
variables
$X_{IK}$ leads to the general form of the
orthogonality relations
\begin{eqnarray}
 \int [dU] {\cal D}_{IJ}^{\{s\}} (U)
{\cal D}_{KL}^{\{t\}*} (U) = (-1)^ {\epsilon^{\{s\}}_J} \alpha_{\{{t}\}}
\delta^{\{s\},\{t\}} \delta_{IK}^{\{s\}} \delta_{JL}^{\{t\}},
\label{ORTO}
\end{eqnarray}
where $(U^\dagger)_{ij}=(U^{-1})_{ij}=(U^{*})_{ji}$. Our notation is such
that the
fundamental representation is labeled with lower case indices $i_1, i_2,
\ldots,i_q$ and capital letter indices denote a family of lower
case indices,
i.e $I=\{i_1, i_2, \ldots,i_p \}$, for example.

In equation (\ref{ORTO}) we
have restricted ourselves to the supergroup $U(m|n)$. Except for the
$(-1)^ {\epsilon^{\{s\}}_J}$ factor that appears as a consequence
of dealing
with Grassmann numbers, the general form of the orthogonality relations
(\ref{ORTO}) does
not apparently  differ from  that of the $U(N)$ case.  However,
as we will
see in the sequel, the
determination of the $\alpha_{\{t\}}$ coefficients will be crucial
in stating
their difference.

\subsection{Null integral over the $U(m|n)$ measure}
\label{ORFU3}

As opposed to what happens in the $U(N)$ case, the determination of the
coefficients $\alpha_{\{t\}}$'s will be much more involved in our
case.  The
reason for this is  the unexpected  normalization condition which is used
for the
determination of these coefficients.  In the $U(N)$ case this
normalization is
$\int [dU] = 1$, while  in our case it  turns out  to be
\begin{eqnarray}
\label{intnull}
\int[dU] = 0, & \mbox{$U \in U(m|n)$}.
\end{eqnarray}

Although this result was known before \cite{berezin1} it emerges
naturally when we deal with
the SUSY
HCIZ
integral (see Appendix \ref{SUSYHCIZ} ).
On one hand, by setting $\beta=0$ on its definition, we directly
obtain the
integral over the measure of the supergroup.  On the other hand,
using its
explicit result we have to calculate $\lim_{ \beta \rightarrow 0}
[\beta^{mn}
I(\lambda_1, \lambda_2, \beta) I(\bar{\lambda}_1, \bar{\lambda}_2,
-\beta)]$,
where $I(\lambda_1, \lambda_2, \beta)$ is the standard HCIZ for
$U(m)$.  Since
$I(\lambda_1, \lambda_2, \beta=0)=1$, we obtain the desired result.

An important application in this work will be the characterization
of the undotted and dotted
representations $ {\{s\}}$ of $U(m|n)$ for which $\alpha_{\{s\}}
\neq 0$.
But before going ahead with the determination of the
$\alpha_{\{t\}}$'s we
briefly show two immediate consequences of (\ref{intnull}).

(i) Choosing a fixed representation $\{s\}$ and summing with
respect to $J=L$
in
equation (\ref{ORTO}) we are left with the constraint
\begin{equation} \int
[dU]=0=\alpha_{\{s\}} str I_{(m_{\{s\}}+n_{\{s\}})\times
(m_{\{s\}}+n_{\{s\}})}.
\label{CONST1} \end{equation} In particular,
this means that all representations
with $\alpha_{\{s\}} \neq 0$
will ne\-ces\-saril\-ly have a null supertrace for the
unit supermatrix in the representation  $\{s\}$.

(ii) From equations (\ref{ORTO}) and (\ref{CONST1}) we obtain
\begin{equation}
\int [dU] s\chi_{\{s\}} (U)s\chi^{*}_{\{t\}} (U)=0,
\end{equation}
even if $\{s\} = \{t\}$,
because this
relation involves again the supertrace of the corresponding  unit
supermatrix.

\subsection{Determination of the $\alpha_{\{t\}}$ coefficient for
ordinary representations}
\label{ORFU4}

If we introduce only one supercharacter in the integration of
Eq.(\ref{ORTO}),
we are left with
\begin{equation}
\int [dU] s\chi_{\{s\}}(U){\cal  D}_{KL}^{\{t\}*}
(U) = \alpha_{\{{t}\}} \delta^{\{s\},\{t\}} \delta^{\{t\}}_{KL},
\label{ORTOC}
\end{equation}
which plays the role of the standard orthogonality condition of the
characters
in the classical case.

The condition (\ref{ORTOC}) implies de following useful

{\it Lemma}:  The supercharacters $s\chi_{\{t\}} (U)\equiv \sum_I
(-1)^{\epsilon^{\{t\}}_I} {\cal D}_{II}^{\{t\}}(U) $ of the
representations
${\cal D}^{\{t\}} (U)$ for which $\alpha_{\{ t \}} \neq 0$
constitute a
linearly
independent set.

The proof goes as follows:  let us consider a null linear combination of
supercharacters of representations with $\alpha_{\{s\}} \neq 0$:
$\sum_{\{s\}}
a_{\{s\}} s\chi_{\{s\}} (U) = 0$.  Multiplying this equation by
${\cal D}_{KL}^{\{t\}*} (U)$, integrating over $[dU]$ and using
Eq.(\ref{ORTOC}) we have
$a_{\{t\}} \alpha_{\{{t} \}} \delta^{\{t\}}_{kl} = 0$ for each
representation
$\{t\}$,
which
shows that $a_{\{t\}} = 0$ provided $\alpha_{\{{t} \}}\neq 0$ .

The starting point that leads to the determination of the undotted
representations $\{t\}$ which have non-zero values for
$\alpha_{\{{t} \}}$ in
(\ref{ORTO}) is the supersymmetric extension of the HCIZ integral
given in
Refs.\cite{Nosotros1, GUHR2}.

A convenient way of rewriting the standard HCIZ integral (defined
in equation
(\ref{ordinary}), Appendix \ref{SUSYHCIZ})
is in terms of its
expansion in characters of the corresponding irreducible
representations of the unitary group \cite{IZ}
\begin{eqnarray}
I(\lambda_1, \lambda_2;\beta) =
\sum_{\{n\}} \frac{\beta^{|n|}}{|n|!}  \frac{\sigma_{\{n\}}}{d_{\{n\}}}
\chi_{\{n\}} (\lambda_1) \chi_{\{n\}} (\lambda_2),
\label{char-IZ}
\end{eqnarray}
where $d_{\{n\}}$ is the dimension of the representation ${\{n\}}$ and
$\sigma_{\{n\}}$ and $|n|$ were already defined in (\ref{SIGMA}) and
(\ref{sigmanueva}).

It will prove convenient for our purposes, to obtain the analogous
supercharacter expansion of the expression given in
(\ref{SUSYIZUBER}) for the
SUSY HCIZ integral.
This we do by using the orthogonality relations (\ref{ORTO}).
The construction goes as follows:  starting
from the SUSY
HCIZ integral
\begin{eqnarray} \tilde{I}(M_1, M_2;\beta) = \int [dU] e^{\beta
str(M_1 U M_2 U^{\dagger})}= \int [dU] \sum_{p=0}^{\infty} \frac
{\beta^p}{p!}
(str(M_1 U M_2 U^{\dagger}))^p,
\label{superIZ}
\end{eqnarray}
and using the
result in (\ref{formula}) we get
\begin{eqnarray}
\tilde{I}(M_1, M_2; \beta) =
\sum_{p=0}^{\infty} \frac {\beta^p}{p!}  {\sum_{\{t\}}}^{'}
\sigma_{\{t\}} \int[dU]
s\chi_{\{t\}} (M_1 U M_2 U^{\dagger}),
\label{superequation}
\end{eqnarray}
where the representations that contribute to the above primed sum are the
ones for
which $|t|=p$, for a given $p$.

Let us now calculate the integral
\begin{eqnarray}
I_{{\{t\}}}(M_1, M_2) = \int
[dU] s\chi_{{\{t\}}} (M_1 U M_2 U^{\dagger}).
\label{B}
\end{eqnarray}
Using the
definition of the supercharacter together with the properties of a
representation we have
\begin{eqnarray}
I_{{\{t\}}}(M_1, M_2) = \int [dU]
\sum_{a=1}^{sd_{\{t\}}} (-1)^{\epsilon_a} {\cal D}_{aa}^{{\{t\}}}
(M_1 U M_2
U^{\dagger}) \nonumber \\
= \int dU \sum_{a,b,c,d=1}^{sd_{\{t\}}}
(-1)^{\epsilon_a} {\cal D}_{ab}^{{\{t\}}} (M_1) {\cal
D}_{bc}^{{\{t\}}} (U)
{\cal D}_{cd}^{{\{t\}}} (M_2) {\cal D}_{da}^{{\{t\}}} (U^{\dagger})
\nonumber \\
= \int [dU] \sum_{a,b,c,d=1}^{sd_{\{t\}}} (-1)^{\epsilon_a}
(-1)^{(\epsilon_b +
\epsilon_c)(\epsilon_c + \epsilon_d)} {\cal D}_{ab}^{{\{t\}}} (M_1) {\cal
D}_{cd}^{{\{t\}}} (M_2) {\cal D}_{bc}^{{\{t\}}} (U) {\cal
D}_{da}^{{\{t\}}}
(U^{\dagger}) \nonumber \\ = \sum_{a,b,c,d=1}^{sd_{\{t\}}}
(-1)^{\epsilon_a}
(-1)^{(\epsilon_b + \epsilon_c)(\epsilon_c + \epsilon_d)} {\cal
D}_{ab}^{{\{t\}}} (M_1) {\cal D}_{cd}^{{\{t\}}} (M_2)
\underbrace{\int dU {\cal
D}_{bc}^{{\{t\}}} (U) {\cal D}_{da}^{{\{t\}}} (U^{\dagger})
}_{(-1)^{\epsilon_c}
\delta_{ba} \delta_{cd} {\alpha}_{{\{t\}}}}
\end{eqnarray}
Finally we obtain
\begin{eqnarray}
I_{\{t\}}(M_1, M_2) = {\alpha}_{\{t\}} \sum_{a,c=1}^{sd_{\{t\}}}
(-1)^{\epsilon_a} (-1)^{\epsilon_c} {\cal D}_{aa}^{\{t\}} {\cal
D}_{cc}^{\{t\}}
\nonumber \\ \Rightarrow I_{\{t\}}(M_1, M_2) = {\alpha}_{\{t\}}
s\chi_{\{t\}}
(M_1) s\chi_{\{t\}} (M_2)
\label{B2}
\end{eqnarray}
Substituting this last result
in equation (\ref{superequation}) we get the expansion in
supercharacters for the SUSY HCIZ integral
\begin{eqnarray}
\tilde{I}(M_1,
M_2;\beta) = \sum_{\{ t \} } \frac{\beta^{|t|}}{|t|!}  \sigma_{\{ t \} }
\alpha_{\{t\}} s\chi_{\{t\}}(M_1) s\chi_{\{t\}}(M_2),
\label{char-SUSY-IZ}
\end{eqnarray}
which contains only undotted representations.

In virtue of the Lemma proved at the begining of this section, we
see that the
representations which contribute to Eq.(\ref{char-SUSY-IZ}) have
supercharacters
that form a linearly independent set.

Up to now, the $\alpha_{\{t\}}$'s are still unknowns. Next  we
identify the
representations with non-zero $\alpha_{\{t\}}$.
 The basic
expression we use is the character expansion in both sides of
Eq.(\ref{eq-import}),
which is
\begin{eqnarray}
\sum_{\{ t\}} \frac{\beta^{|t|}}{|t|!}
\sigma_{\{t\}}
\alpha_{\{t\}} s\chi_{\{t\}}(M_1) s\chi_{\{t\}}(M_2) = \sum_{\{p\}}
\sum_{\{q\}}
\frac{\beta^{|p|+|q|+mn}}{|p|!  |q|!}  \frac{\sigma_{\{p\}}
\sigma_{\{q\}} }
{d_{\{p\}} d_{\{q\}}} (-1)^{|q|} \times \nonumber \\ \times
\Sigma(\lambda_1,\bar{\lambda}_1) \chi_{\{p\}} (\lambda_1) \chi_{\{q\}}
(\bar{\lambda}_1) \Sigma(\lambda_2, \bar{\lambda}_2) \chi_{\{p\}}
(\lambda_2)
\chi_{\{q\}} (\bar{\lambda}_2), \
\label{main}
\end{eqnarray}
where
\begin{eqnarray}
\Sigma (\lambda,\bar\lambda) =
\prod_{i=1}^m \prod_{\alpha=1}^n
( \lambda_i - {\bar\lambda}_{\alpha} ).
\label{Sigma2}
\end{eqnarray}
Now we analize this
equation by considering the following cases:  \subsubsection { Case
of $|t|<
mn$} Before making any further analysis, from (\ref{main}) we can
immediately
conclude that
\begin{equation}
\alpha_{\{t\}} = 0, \ \mbox{for} \ |t|=
0,1,\ldots ,(mn-1).
\label{TMENOR}
\end{equation}
This is because in
both sides
of that equation we have a power series in $\beta$, and the right
hand side
(RHS) of it starts with $\beta^{mn}$ while the left hand side (LHS)
starts with
$\beta^0$.  The proof goes by assuming that some coefficients
$\alpha_{\{t\}}$
are non-zero.  The linear independence of the $s\chi_{\{t\}}(M)$'s
associated
to
those representations imply that $\alpha_{\{t\}}$ must be zero.

\subsubsection { Case of $|t| \geq mn$}

As we just  said  before,
Eq.(\ref{main})
is a power series in $\beta$ , so for
a given power $|t|$ of  $\beta$ we obtain
\begin{eqnarray}
\frac{1}{|t|!}
{\sum_{\{t\}}}^{'} \sigma_{\{t\}} \alpha_{\{t\}} s\chi_{\{t\}}(M_1)
s\chi_{\{t\}}(M_2) = \sum_{\{p\}} \sum_{\{q\}}
\frac{(-1)^{|q|}}{|p|!  |q|!}
\frac{\sigma_{\{p\}} \sigma_{\{q\}} }{d_{\{p\}} d_{\{q\}}} \times \
\ \ \ \ \ \
\ \ \nonumber \\ \times \Sigma(\lambda_1,\bar{\lambda}_1) \chi_{\{p\}}
(\lambda_1) \chi_{\{q\}} (\bar{\lambda}_1) \ \Sigma(\lambda_2,
\bar{\lambda}_2)
\chi_{\{p\}} (\lambda_2) \chi_{\{q\}} (\bar{\lambda}_2),
\label{power}
\end{eqnarray}
where the sum in the LHS is made for all tableaux
having a fixed
number of boxes $|t|$, while the sum over $\{p\}$ and $\{q\}$ in
the RHS is
restricted to \begin{equation} |p|+|q| = |t| - mn.  \label{restriction}
\end{equation}
We now want to prove that Eq.(\ref{power})
necessarily implies
that
\begin{eqnarray}
s\chi_{\{t\}}(M) = c^{\{t\}}_{\{p\},\{q\}}
\Sigma(\lambda,\bar{\lambda}) \chi_{\{p\}} (\lambda)
\chi_{\{q\}} (\bar{\lambda}),
\label{want}
\end{eqnarray}
for some $\{p\}$ and $\{q\}$ satisfying
(\ref{restriction}) and
for a certain representation $\{t\}$ that we will determine.

In order to extract more information from Eq.(\ref{power}) let us
consider an
arbitrary supermatrix $M_2$, while we restrict the supermatrix
$M_1=\tilde{M}$
in such a way that one of its $\lambda$-eigenvalues be equal to one
of its
$\bar{\lambda}$- eigenvalues.  Namely, let
$\lambda_j=\bar{\lambda}_{\beta}$,
for
example.  Then, in Eq.(\ref{power}) we are left with
\begin{eqnarray}
\frac{1}{|t|!}  \sum_{\{t\}} \sigma_{\{t\}} \alpha_{\{t\}}
s\chi_{\{t\}}(\tilde{M}) s\chi_{\{t\}}(M_2) = 0, \label{null}
\end{eqnarray}
because $\Sigma(\lambda_1,\bar{\lambda}_1)$ becomes zero.  If we
look at this
relation as a null linear combination of the supercharacters
$s\chi_{\{t\}}(M_2)$ with coefficients
\begin{eqnarray} \gamma_{\{t\}} =
\frac{1}{|t|!}  \sigma_{\{t\}} \alpha_{\{t\}} s\chi_{\{t\}}(\tilde{M}),
\label{gamma}
\end{eqnarray}
we conclude that the coefficients
$\gamma_{\{t\}}$
are all zero, because the supercharacters appearing in (\ref{null})
constitute
a
linearly independent set.  But $\sigma_{\{t\}}$ and $\alpha_{\{t\}}$ are
different from zero, so that we are left with
$s\chi_{\{t\}}(\tilde{M})=0$.
Recalling that $s\chi_{\{t\}}(M)$ is a polynomial function of the
$\lambda_i$'s and the $\bar{\lambda}_{\alpha}$'s, we conclude from this
relation
that $s\chi_{\{t\}}(M)$ must be divisible by
$(\lambda_j - \bar{\lambda}_{\beta})$. That is to say
\begin{eqnarray}
s\chi_{\{t\}}(M) = (\lambda_j - \bar{\lambda}_{\beta})
F_{j\beta}(\lambda,\bar{\lambda}), \label{F}
\end{eqnarray}
where $F_{j\beta}(\lambda,\bar{\lambda})$ is another polynomial
function of the
eigenvalues.
The same reasoning can be extended to every $\lambda_i \ (i=1,...,m)$ and
$\bar{\lambda}_{\alpha} \ (\alpha = 1,...,n)$, and this implies that
$s\chi_{\{t\}}(M)$ must have the form
\begin{eqnarray}
s\chi_{\{t\}}(M) = \prod_{i=1}^m \prod_{\alpha =1}^n
(\lambda_i -\bar{\lambda}_{\alpha}) \ P(\lambda, \bar{\lambda})=
\Sigma(\lambda,\bar{\lambda}) \ P(\lambda, \bar{\lambda}).
\label{Sigma1}
\end{eqnarray}
In Eq.(\ref{Sigma1}), $P(\lambda,\bar{\lambda})$ must be an
homogeneous polynomial
function of all the eigenvalues, because $s\chi_{\{t\}}(\tilde{M})$ and
$\Sigma(\lambda, \bar{\lambda})$ are so.  The degree of homogeneity of
$s\chi_{\{t\}}(\tilde{M})$ and $\Sigma(\lambda, \bar{\lambda})$ is
$|t|$ and
$mn$, respectively.  This means that the degree of homogeneity of
$P(\lambda,\bar{\lambda})$ must be $|t|-mn$.  Also, we know that
$s\chi_{\{t\}}(\tilde{M})$ and $\Sigma(\lambda, \bar{\lambda})$ are
symmetric
functions in the eigenvalues $\lambda_i$, $\bar{\lambda}_{\alpha}$,
separately,
and so should be $P(\lambda,\bar{\lambda})$.  Summing up then,
$P(\lambda,\bar{\lambda})$ is:  (i) an homogeneous polynomial function of
degree
$|t|-mn$ in all the eigenvalues and (ii) a symmetric function of
the $\lambda_i$'s
and the $\bar{\lambda}_{\alpha}$'s, separately.  Since the characters
$\chi_{\{a\}}(\lambda)$ \ $\left(\chi_{\{b\}}(\bar{\lambda}) \right)$ are
polynomial homogeneous functions of degree $|a|$ \ $(|b|)$, which
are symmetric
in the eigenvalues $\lambda_i$ \ $(\bar{\lambda}_\alpha)$ and constitute
a complete
linearly independent set, $P(\lambda,\bar{\lambda})$ can be written as
\begin{eqnarray} P(\lambda,\bar{\lambda}) = \sum_{\{a\},\{b\}}
c_{\{a\},\{b\}}^{\{t\}} \chi_{\{a\}}(\lambda) \chi_{\{b\}}
(\bar{\lambda}),
\label{P} \end{eqnarray} where the sum in $\{a\}$ and $\{b\}$ is
rectricted by
$
|a|+|b|=|t|-mn$.  Substituing this last relation in (\ref{Sigma1})
we have
\begin{eqnarray}
s\chi_{\{t\}} (M) = \Sigma (\lambda, \bar{\lambda})
\sum_{\{a\},\{b\}} c_{\{a\},\{b\}}^{\{t\}} \chi_{\{a\}}(\lambda)
\chi_{\{b\}}
(\bar{\lambda}).
\label{new1}
\end{eqnarray}
Using the above expression in the
LHS of (\ref{power}) and comparing both sides of this equation, we
conclude
that
the  RHS of  (\ref{new1}) should be saturated only with one
coefficient, for
a certain
tableaux ${\{t\}}$, which precise form is yet to be determined.  That is
\begin{eqnarray*}
s\chi_{\{t\}} (M) = c^{{\{t\}}}_{\{p\},\{q\}}
\Sigma (\lambda,
\bar{\lambda}) \chi_{\{p\}}(\lambda) \chi_{\{q\}} (\bar{\lambda}),
\end{eqnarray*}
where $\{p\}$ and $\{q\}$ satisfy (\ref{restriction}). Thus, we
have proved
our result  in (\ref{want}).

In order to identify the Young tableau corresponding to the
representation
${\{t\}}$ we will make use of the fact that the tableaux structure is
independent of whether we are dealing with a group or supergroup.
Of course,
the specific symmetrization ( antisimmetrization) properties will
be different
in each case.  In this way we will identify the tableaux by looking
only at the
known characters of the $U(m), U(n)$ subgroups of $U(m|n)$, in
Eq.(\ref{want}).

\subsubsection*{4.4.2.1 The case of $\{p\} = \{q\} = 0$ }

Here we
have $|t|=mn$ and \begin{equation} s\chi_{\{t\}} (M) = c^{
{\{t\}}}_{\{0\},\{0\}} \Sigma (\lambda, \bar{\lambda}).  \label{c00}
\end{equation} In order to proceed with the required
identifications, let us
consider the particular case where the only non-zero block of the
supermatrix
$M$ is the $m\times m$ block, i.e.  \begin{eqnarray} M = \left (
\begin{array}{cc} M' & 0 \\ 0 & 0 \end{array} \right ).  \label{case1}
\end{eqnarray} Then Eq.(\ref{c00}) reduces to \begin{eqnarray}
\chi_{\{t\}}
(M')
= c^{{\{t\}}}_{\{0\},\{0\}} (\prod_{i=1}^m \lambda_i )^n.  \label{c00M'}
\end{eqnarray} Using Weyl's formula for the character of the
representations of
the unitary group \cite{Weyl}
\begin{eqnarray} \chi_{\{r\}} (\lambda) = \frac {det
(\lambda_i{}^{r_j +n -j})} {det (\lambda_i{}^{n-j})}
\label{weyl}
\end{eqnarray}
we conclude that the product of eigenvalues in (\ref{c00M'})
corresponds to the
character of the representation $\{r\}=(r_1,r_2,\ldots,r_m)$ with
$r_i=n$ of
$U(m)$, which we denote by $\{r\} =\{mn\}$. So,  pictorically,
$\{r\}$ will look like
\begin{center}
\begin{picture}(36,84)(0,0)
\put(0,0){\framebox(12,12){$\space$}}
\put(12,0){\framebox(12,12){$\space$}}
\put(24,0){\framebox(12,12){$\space$}}
\put(0,12){\framebox(12,12){$\space$}}
\put(12,12){\framebox(12,12){$\space$}}
\put(24,12){\framebox(12,12){$\space$}}
\put(-96,27){$\{r\} =\{mn\}=$} \put(-12,27){$m$}
\put(0,24){\framebox(12,12){$\space$}}
\put(12,24){\framebox(12,12){$\space$}}
\put(24,24){\framebox(12,12){$\space$}}
\put(0,36){\framebox(12,12){$\space$}}
\put(12,36){\framebox(12,12){$\space$}}
\put(24,36){\framebox(12,12){$\space$}}
\put(0,48){\framebox(12,12){$\space$}}
\put(12,48){\framebox(12,12){$\space$}}
\put(24,48){\framebox(12,12){$\space$}} \put(15,63){$n$}
\end{picture}
\end{center}
\begin{eqnarray}
\label{tableaumn}
\end{eqnarray}
In this way we have that $\chi_{\{t\}} (M') = c_{\{0\},\{0\}}^{\{t\}}
\chi_{(n,n, \ldots ,n)} (M')$, which allows the identification of the
representation $\{ t \}$ as the one given by the tableau
corresponding to $
t_1=t_2= \ldots = t_m = n$, pictorically shown in (\ref{tableaumn}),
together with $c^{\{t\}}_{\{0\},\{0\}} = 1$. We will denote by
$\{mn\}$ the
representation just found.
Besides, we identify $\Sigma (\lambda, \bar{\lambda})$ as the
supercharacter of
the representation referred to above:
\begin{eqnarray}
s \chi_{\begin{picture}(12,28)(0,0)
\put(0,0){\framebox(4,4){$\space$}}
\put(4,0){\framebox(4,4){$\space$}}
\put(8,0){\framebox(4,4){$\space$}}
\put(0,4){\framebox(4,4){$\space$}}
\put(4,4){\framebox(4,4){$\space$}}
\put(8,4){\framebox(4,4){$\space$}}
\put(0,8){\framebox(4,4){$\space$}}
\put(4,8){\framebox(4,4){$\space$}}
\put(8,8){\framebox(4,4){$\space$}}
\put(0,12){\framebox(4,4){$\space$}}
\put(4,12){\framebox(4,4){$\space$}}
\put(8,12){\framebox(4,4){$\space$}}
\put(0,16){\framebox(4,4){$\space$}}
\put(4,16){\framebox(4,4){$\space$}}
\put(8,16){\framebox(4,4){$\space$}}
\end{picture}} (M) = \Sigma(\lambda, \bar{\lambda}),
\label{nontrivial}
\end{eqnarray}
where $\Sigma(\lambda, \bar{\lambda})$ is given in (\ref{Sigma2}).


\subsubsection*{4.4.2.2 The case of $\{p\} \neq 0$, $\{q\}=0$}


Here we
have $|t|=|p|+mn$ and $s\chi_{\{t\}} (M)= c^{\{ t \}}_{\{p\},\{0\}}
\Sigma
(\lambda, \bar{\lambda}) \chi_{\{p\}} (\lambda)$.  Considering in this
expression the same choice of $M$ as in (\ref{case1}), we have
$\chi_{\{t\}}
(M') = c^{\{t\}}_{\{p\},\{0\}} (\prod_{i=1}^m \lambda_i )^n \chi_{\{p\}}
(\lambda)$.  Using again Weyl's formula we are able to make the
identification
$(\prod_{i=1}^m \lambda_i )^n \chi_{\{p\}} (\lambda) = \chi_{\{n+p\}}
(\lambda)$, where by ${\{n+p\}}$ we mean the representation with
Young tableau
$
(n+p_1, n+p_2, \ldots , n+p_m )$:
\begin{center}
\begin{picture}(90,84)(0,0)
\put(0,0){\framebox(12,12){$\space$}}
\put(12,0){\framebox(12,12){$\space$}}
\put(24,0){\framebox(12,12){$\space$}}
\put(0,12){\framebox(12,12){$\space$}}
\put(12,12){\framebox(12,12){$\space$}}
\put(24,12){\framebox(12,12){$\space$}}
\put(40,12){\framebox(12,12){$\space$}}
\put(-108,27){$\{r\} =\{mn\}\{p\}=$}
\put(-12,27){$m$}
\put(0,24){\framebox(12,12){$\space$}}
\put(12,24){\framebox(12,12){$\space$}}
\put(24,24){\framebox(12,12){$\space$}}
\put(40,24){\framebox(12,12){$\space$}}
\put(0,36){\framebox(12,12){$\space$}}
\put(12,36){\framebox(12,12){$\space$}}
\put(24,36){\framebox(12,12){$\space$}}
\put(40,36){\framebox(12,12){$\space$}}
\put(52,36){\framebox(12,12){$\space$}}
\put(64,36){\framebox(12,12){$\space$}}
\put(0,48){\framebox(12,12){$\space$}}
\put(12,48){\framebox(12,12){$\space$}}
\put(24,48){\framebox(12,12){$\space$}}
\put(40,48){\framebox(12,12){$\space$}}
\put(52,48){\framebox(12,12){$\space$}}
\put(64,48){\framebox(12,12){$\space$}}
\put(76,48){\framebox(12,12){$\space$}}
\put(15,63){$n$}
\end{picture}
\end{center}
\begin{eqnarray}
\label{mnp-tableau}
\end{eqnarray}
where we have generically drawn

\begin{center}
\begin{picture}(48,48)(0,0)
\put(-36,21){$\{p\}=$}
\put(0,0){\framebox(12,12){$\space$}}
\put(0,12){\framebox(12,12){$\space$}}
\put(0,24){\framebox(12,12){$\space$}}
\put(12,24){\framebox(12,12){$\space$}}
\put(24,24){\framebox(12,12){$\space$}}
\put(0,36){\framebox(12,12){$\space$}}
\put(12,36){\framebox(12,12){$\space$}}
\put(24,36){\framebox(12,12){$\space$}}
\put(36,36){\framebox(12,12){$\space$}}
\end{picture}
\end{center}
\begin{eqnarray}
\label{p-tableau}
\end{eqnarray}

This leads to $\chi_{\{t\}} (M') = c^{\{t\}}_{\{p\},\{0\}} \chi_{(n+p_1,
n+p_2, \ldots
, n+p_m )} (\lambda)$ for this case and we conclude that
$c^{\{t\}}_{\{p\},\{0\}} = 1$ with ${\{t\}}$ being the
representation $( n+p_1,
n+p_2, \dots , n+p_m )$ of $U(m|n)$.  Besides, we identify
\begin{eqnarray}
s\chi_{\{t\}} (M)=
\Sigma (\lambda,\bar{\lambda}) \chi_{\{p\}} (\lambda)
\label{schip}
\end{eqnarray}

\subsubsection*{ 4.4.2.3 The case of arbitrary $\{p\}$ and $\{q\}$}

Now we
discuss the main result of this section which states that the undotted
representations of
$U(m|n)$ with $\alpha_{\{t\}}\neq0$ are characterized by the
following Young
tableaux:
\begin{center}
\begin{picture}(116,84)(0,0)
\put(0,-16){\framebox(12,12){$\space$}}
\put(12,-16){\framebox(12,12){$\space$}}
\put(0,-28){\framebox(12,12){$\space$}}
\put(12,-28){\framebox(12,12){$\space$}}
\put(0,-40){\framebox(12,12){$\space$}}
\put(0,0){\framebox(12,12){$\space$}}
\put(12,0){\framebox(12,12){$\space$}}
\put(24,0){\framebox(12,12){$\space$}}
\put(0,12){\framebox(12,12){$\space$}}
\put(12,12){\framebox(12,12){$\space$}}
\put(24,12){\framebox(12,12){$\space$}}
\put(40,12){\framebox(12,12){$\space$}}
\put(-108,27){$\{\tilde{t}\} =\{mn\}\{p\}=$}
\put(-76,11){$\{q\}^T$}
\put(-12,27){$m$} \put(0,24){\framebox(12,12){$\space$}}
\put(12,24){\framebox(12,12){$\space$}}
\put(24,24){\framebox(12,12){$\space$}}
\put(40,24){\framebox(12,12){$\space$}}
\put(0,36){\framebox(12,12){$\space$}}
\put(12,36){\framebox(12,12){$\space$}}
\put(24,36){\framebox(12,12){$\space$}}
\put(40,36){\framebox(12,12){$\space$}}
\put(52,36){\framebox(12,12){$\space$}}
\put(64,36){\framebox(12,12){$\space$}}
\put(0,48){\framebox(12,12){$\space$}}
\put(12,48){\framebox(12,12){$\space$}}
\put(24,48){\framebox(12,12){$\space$}}
\put(40,48){\framebox(12,12){$\space$}}
\put(52,48){\framebox(12,12){$\space$}}
\put(64,48){\framebox(12,12){$\space$}}
\put(76,48){\framebox(12,12){$\space$}}
\put(15,63){$n$} \put(0,-16){\framebox(12,12){$\space$}}
\put(12,-16){\framebox(12,12){$\space$}}
\put(0,-28){\framebox(12,12){$\space$}}
\put(12,-28){\framebox(12,12){$\space$}}
\put(0,-40){\framebox(12,12){$\space$}}
\end{picture}
\end{center}
\begin{eqnarray}
\label{main-tableau}
\end{eqnarray}
\noindent where $\{p\}$ is the same as in (\ref{p-tableau}) and
$\{q\}$ is pictorically identified with
\begin{center}
\begin{picture}(36,24)(0,0)
\put(-36,9){$\{q\}=$}
\put(0,0){\framebox(12,12){$\space$}}
\put(12,0){\framebox(12,12){$\space$}}
\put(0,12){\framebox(12,12){$\space$}}
\put(12,12){\framebox(12,12){$\space$}}
\put(24,12){\framebox(12,12){$\space$}}
\end{picture}
\end{center}
\begin{eqnarray}
\label{q-tableau}
\end{eqnarray}
\noindent
which, after interchanging its rows and columns, giving $\{q\}^T$,
is put in
the
bottom left of $\{mn\}\{p\}$ , producing (\ref{main-tableau}).  The
representation $\{q\}^T$ is called the conjugate representation of
$\{q\}$.

Besides identifying the particular representations involved, we are
also able
to
calculate the corresponding non-zero normalization coefficient
appearing in the
orthogonality relations (\ref{ORTO}) for the representation
$\{\tilde{t}\}$.
It is given by
\begin{eqnarray}
\alpha_{\{{\tilde t}\}} = (-1)^{|q|} \frac{ |{\tilde t}|!}{|p|!  |q|!}
\frac{\sigma_{\{p\}} \sigma_{\{q\}} }{\sigma_{\{{\tilde t}\}} }
\frac{1}{d_{\{p\}} d_{\{q\}}}.
\label{finalresult}
\end{eqnarray}

 Let us also remark
that our expression (\ref{finalresult}) correctly reproduces the result
\begin{equation}
\alpha_{\{t\}}={1\over d_{\{t\}}}
\label{1sobred}
\end{equation}
for $U(N)$ (by making $m=N$ and $n=0$).
Note that  in the $U(m|n)$
case, the $\alpha_{\{t\}}$ coefficient not only depends on the
dimension of the
representations involved, but also on the
Clebsch-Gordan coefficients $\sigma_{\{t\}}$ and on the
characteristic number
$|t|$.

An important result that leads to the above conclusions is that
\begin{eqnarray}
s\chi_{\{{\tilde t}\}} (M) = (-1)^{|q|} \Sigma (\lambda, \bar{\lambda})
\chi_{\{p\}}(\lambda) \chi_{\{q\}} (\bar{\lambda}).  \label{resultcpq}
\end{eqnarray}

This relation is proved in Appendix  \ref{chi},  and after
substituting it in
equation (\ref{main}), the result in (\ref{finalresult}) is
obtained.

None of the results (\ref{nontrivial}), (\ref{schip}) and
(\ref{resultcpq})
seem to be easily proved by standard
methods like the supercharacter general formula (\ref{supercar-formula})
of Appendix
\ref{charactertable} or the determinant formulas of
Ref. \cite{BARS2}.
This last formula consists in calculating the determinant of a
matrix which
components are supercharacters of completely symmetric representations.
These supercharacters are expressed in terms of sums
which, apparently, cannot be cast in closed form. Thus, our method
provides
an alternative  derivation of the  compact results already mentioned.

An immediate consequence of our relation  (\ref{resultcpq}) is
that we can
obtain the dimension $sd_{\{t\}}$ for the representations in
$U(m|n)$ that
arise
in the
supercharacter expansion,   in terms of the dimension
$d_{\{p\}}$ ($d_{\{q\}}$) of the
 $U(m)$ ( $U(n)$) representations.  Taking
\begin{eqnarray}
M_0 = \left( \begin{array}{rr} I_{m \times m} & 0 \\ 0 & -I_{n
\times n} \end{array} \right)
\label{mat-dim}
\end{eqnarray}
in (\ref{resultcpq}) and observing
that \cite{BARS2}
\begin{eqnarray*}
s\chi_{\{t\}} (M_0) \rightarrow  sd_{\{t\}} \\
\chi_{\{p\}}(\lambda) \rightarrow d_{\{p\}} \\
\chi_{\{q\}}(-\bar{\lambda}) \rightarrow (-1)^{|q|} d_{\{q\}} \\
\Sigma(\lambda, \bar{\lambda}) \rightarrow 2^{mn},
\end{eqnarray*}
we obtain
the closed expression
\begin{eqnarray}
sd_{\{{\tilde t}\}}= 2^{mn} d_{\{p\}} d_{\{q\}},
\label{superdimension}
\end{eqnarray}
for the dimensions of the representations
of $U(m|n)$ characterized by the tableaux in (\ref{main-tableau}).

Again, it  should be possible to  derive  the general  expression
(\ref{superdimension}) for the dimension
of the general tableaux  (\ref{main-tableau}) by  using the formula
developped by Balantekin and Bars \cite{BARS2} as a determinant of the
supercharacters
of completely symmetric representations. Nevertheless,   we
have not been able to reproduce the general result
(\ref{superdimension}) in this way.

Before closing this section, let us illustrate the formula
(\ref{resultcpq}).
Consider the representation
\begin{eqnarray}
\begin{picture}(36,24)(0,0)
\put(-36,9){$\{\tilde{t}\}=$}
\put(0,0){\framebox(12,12){$\space$}}
\put(12,0){\framebox(12,12){$\space$}}
\put(0,12){\framebox(12,12){$\space$}}
\put(12,12){\framebox(12,12){$\space$}}
\put(24,12){\framebox(12,12){$\space$} \ ,}
\end{picture}
\label{example2}
\end{eqnarray}
whose supercharacter can be obtained with the aid of the character
table of
the symmetric group $S_5$ and the general
expression (\ref{supercar-formula}), giving
\begin{eqnarray}
s\chi_{\{\tilde{t}\}}(M) = \frac{1}{24} [ (strM)^5 + 2 (strM)^3 strM^2
- 4(strM)^2 strM^3 & \nonumber \\- 6 strM strM^4
+ 3 strM (strM^2)^2 + 4 strM^2 strM^3].&
\label{example3}
\end{eqnarray}
Let us consider the tableau  (\ref{example2}) as labeling a  $U(1|2)$
representation. So
according
our notation (\ref{main-tableau}) we have that
\begin{center}
\begin{picture}(12,12)(0,0)
\put(-36,6){$\{p\}=$}
\put(0,0){\framebox(12,12){$\space$}}
\end{picture}
\end{center}
\begin{center}
\begin{picture}(12,24)(0,0)
\put(-36,9){$\{q\}=$}
\put(0,0){\framebox(12,12){$\space$} \ .}
\put(0,12){\framebox(12,12){$\space$}}
\end{picture}
\end{center}
Substituting the  supermatrix $M$ in its diagonal form,
\begin{eqnarray}
M = \left ( \begin{array}{ccc}
            \lambda_1 & 0               & 0 \\
            0         & \bar{\lambda}_1 & 0 \\
            0         & 0               & \bar{\lambda}_2
            \end{array}
    \right),
\end{eqnarray}
in the supercharacter expression (\ref{example3}) and after some algebra,
we obtain
\begin{eqnarray}
s\chi_{\{\tilde{t}\}}(M) =[(\lambda_1 - \bar{\lambda}_1)(\lambda_1 -
\bar{\lambda}_2)] \ (\lambda_1) \ (\bar{\lambda}_1 \bar{\lambda}_2),
\end{eqnarray}
which, for $U(1|2)$, can be equivalently written as
\begin{eqnarray}
s\chi_{\{\tilde{t}\}}(M) = (-1)^2 \Sigma (\lambda, \bar{\lambda})
\chi_{\begin{picture}(12,12)(0,0)
\put(0,0){\framebox(6,6){$\space$}}
\end{picture} } (\lambda)
\chi_{\begin{picture}(6,16)(0,0)
\put(0,0){\framebox(6,6){$\space$}}
\put(0,6){\framebox(6,6){$\space$}}
\end{picture} } (\bar{\lambda}).
\label{example4}
\end{eqnarray}
Here $\chi_{\begin{picture}(12,12)(0,0)
\put(0,0){\framebox(6,6){$\space$}}
\end{picture} } (\lambda)$ and
$\chi_{\begin{picture}(6,16)(0,0)
\put(0,0){\framebox(6,6){$\space$}}
\put(0,6){\framebox(6,6){$\space$}}
\end{picture} } (\bar{\lambda})$
are the $U(1)$ and  the $U(2)$ characters of the corresponding
representations,
which
can be taken from the character table in Appendix
\ref{charactertable} and
$\Sigma(\lambda, \bar{\lambda}) = (\lambda_1 - \bar{\lambda}_1)
(\lambda_1 - \bar{\lambda}_2)$. We see,
then, that
(\ref{example4}) is in accordance with our general result
(\ref{resultcpq}).

\section{Determination of
$\alpha_{\{t\}}$ for complex conjugate and mixed
representations}
\label{ACM}

\subsection{The case of complex conjugated (dotted)
representations}
\label{ACM1}
We will
need the following properties of $\bar U$
\begin{equation}
\left( str {\bar U} \right)^p=((str U)^*)^p,
\qquad str\left( {\bar U}^p \right)=
\left( str U^p \right)^*,
\label{tilde}
\end{equation}
which are just a consequence of the
definition of $\bar U$ together with the group property of the
$\ {\bar {}}$
operation.  Since the supercharacter corresponding to the representation
$\{\dot
p \}$ has the same expression as the one corresponding to the
representation
$\{p \}$ except that $U$ is replaced by $\bar U$, the properties
(\ref{tilde})
imply
\begin{equation} s\chi_{\{\dot p \}}(U)=
s\chi_{\{p\}}^*(U)= s\chi_{\{p\}}(U^\dagger),
\label{pdot}
\end{equation}
where the Young tableau of the representation
$\{\dot p \}$ is
the same as that of the representation $\{p \}$ except that all boxes are
dotted.

We will prove that
\begin{eqnarray}
\alpha_{\{\dot t \}} = \alpha_{\{t\}}.
\label{dotted}
\end{eqnarray}
For this purpose we will look for two equivalent expressions for
the integral
\begin{eqnarray}
I_{\{n\}}(M_1, M_2) = \int [dU] s\chi_{\{n\}}(M_1 U M_2 U^\dagger)),
\label{In}
\end{eqnarray}
which we already presented in (\ref{B}) and where $M_1$ and $M_2$ are
hermitian supermatrices.

The first expression is equation (\ref{B2}), namely,
\begin{eqnarray*}
I_{\{n\}}(M_1, M_2) = {\alpha}_{\{n\}} s\chi_{\{n\}}(M_1)
s\chi_{\{n\}} (M_2).
\end{eqnarray*}

Before going to our second
way of calculating (\ref{In})  we  observe that
\begin{eqnarray*}
s\chi_{\{n\}}((M_1 U M_2 U^\dagger)^\dagger) =
s\chi_{\{n\}}(U M_2 U^\dagger M_1)  = s\chi_{\{n\}}(M_1 U M_2 U^\dagger),
\end{eqnarray*}
which implies that for $B=M_1 U M_2 U^\dagger$ we have
$s\chi_{\{n\}}(B^\dagger)=s\chi_{\{n\}}(B)$.

So, using (\ref{pdot}) we have that
$s\chi_{\{n\}}(M_1 U M_2 U^\dagger) = s\chi_{\{\dot{n}\}}(M_1 U M_2
U^\dagger)$
and therefore
\begin{eqnarray}
I_{\{n\}}(M_1, M_2) = I_{\{\dot n\}}(M_1, M_2)=
{\alpha}_{\{\dot n\}} s\chi_{\{\dot n\}}(M_1) s\chi_{\{\dot n\}} (M_2).
\end{eqnarray}
But for a hermitian supermatrix $M$ we have that
$s\chi_{\{\dot{n}\}}(M) = s\chi_{\{n\}}(M)$, and therefore
\begin{eqnarray}
I_{\{n\}}(M_1, M_2)=
{\alpha}_{\{\dot n\}} s\chi_{\{n\}}(M_1) s\chi_{\{n\}} (M_2).
\label{estrella}
\end{eqnarray}
Comparisson of (\ref{estrella}) and (\ref{B2}) leads to our desired
result
in (\ref{dotted}).

\subsection{The case of mixed representations}
\label{ACM2}

We now prove
the following expression for the  $\alpha$-coefficients for the mixed
representations
 in the orthogonality
relations:
\begin{eqnarray}
\alpha_{\{\dot p\} | \{q\}} = \left[ \frac{ |p|!  \
|q|!}{(|p|+|q|)!}  \right] ^2 \left[\frac{1}{\sigma_{\{p\}}
\sigma_{\{q \}}}
\right] ^2 \sum_{\{t\}}^{}{'} \
\rho_{\{t\}}^{\{p\},\{q\}} \ \sigma_{\{t\}}^2 \alpha_{\{t\}}.
\label{alphamix}
\end{eqnarray}
Here the
 $\rho_{\{t\}}^{\{p\},\{q\}}$'s are the Clebsch-Gordan coefficients
which appear in
the
decomposition of the tensor
product of representations ${\{p\}}$ and ${\{q\}}$
\begin{eqnarray}
\{p\}\otimes \{q\} = \oplus_{\{t\}}^{}{'} \
\rho_{\{t\}}^{\{p\},\{q\}} \ \{t\}.
\label{ct}
\end{eqnarray}
They are obtained by applying the Young tableaux rules for multiplying
irreducible representations \cite{Barut}.
Our notation, $\sum^{}{'}_{\{t\}}$ and $\oplus_{\{t\}}^{}{'}$, means that
the sums are carried
only over
the representations satisfying $|t|=|p|+|q|$.

Let us emphasize that all ingredients in our formula
 (\ref{alphamix})  are  known :  the $\alpha_{\{t\}}$'s are either
null or
given by
 (\ref{finalresult}),  the $\rho_{\{t\}}^{\{p\},\{q\}}$'s are given by
(\ref{ct})
and the $\sigma_{\{ p \}}$'s
are given by (\ref{sigmanueva}).  Some
examples of
 the relations (\ref{alphamix}) are given in Appendix  \ref{Alfas}.

 Now, to
 prove (\ref{alphamix}) let us  consider
 \begin{eqnarray}
I_{\{p\},\{q\}}(M_1, M_2) = \int [dU]
 s\chi_{\{\dot p\} | \{q\}} (M_1 U M_2 U^{\dagger})
\label{Ipq}
\end{eqnarray}
and, following the idea of the previous cases, we are going to
calculate this
expression  in two different ways. The method that  will be
subsequently used
consists basically in
comparing  these two expressions  as polynomial expansions
in
$( str M_1^{k_1})^{l_1} ( str M_2^{k_2})^{l_2}$.  For our purposes
it will be
enough
only to consider the highest power term
\begin{eqnarray*}
(str \ M_1)^{|p|+|q|} (str \ M_2)^{|p|+|q|}.
\end{eqnarray*}

Since our argument is based only in the comparison of the highest
power term
$(str \ A)^{|p|+|q|} $ in the corresponding expressions, we
next present the relevant approximations that will produce such
terms. To begin with
we consider the expansion
\begin{eqnarray}
 s\chi_{\{\dot a\} | \{b\}} (A) = s\chi_{\{\dot a\}} (A)
s\chi_{\{b\}} (A) +
\dots,
 \label{term1}
 \end{eqnarray}
 which  complete expression can be found  in  the Appendix
\ref{smixrep}. This
is a
function of supercharacters
 $s\chi_{\{\dot r\}} (A)$ and $s\chi_{\{ s\}} (A)$ (with $|r| \leq
|a|$, $|s|
 \leq |b|$).    For our purposes
it is enough only  to consider  the term written in (\ref{term1}). The
remaining terms will contain
 the factor $ (s tr  (A A^\dagger)^i )^{r_i}$, thus lowering the power of
$(s tr A)$.
The next step is to
express the corresponding  supercharacters in terms of powers of
supertraces.
Again, what we
need is to consider  the highest power term
\begin{eqnarray} s\chi_{\{ n\}} (A)
 =  \frac{\sigma_{\{n\}}}{|n|!}  (str \ A)^{|n|} + \dots.  \label{term2}
 \end{eqnarray}
of the full  polynomial expresion (\ref{supercar-formula}).

In this way, using  (\ref{term1}) and
 (\ref{term2}) for  the case of a hermitian supermatrix M, we have that
 \begin{eqnarray} s\chi_{\{\dot a\} | \{b\}} (M) = \frac{\sigma_{\{a\}}
 \sigma_{\{b\}}}{|a|!  \ |b|!}  (str \ M)^{|a|+|b|} + \dots ,
\label{term3}
 \end{eqnarray}
where we have displayed only the highest power term in $(s tr M)$.
We emphasize  that the coefficient of $(str \
 M)^{|a|+|b|}$ written in this last relation is exact.

 With the above considerations we now proceed with the calculation.
The direct integration over the
 supergroup in Eq.(\ref{Ipq})  gives
\begin{eqnarray}
I_{\{p\},\{q\}}(M_1,M_2) = \alpha_{\{\dot p\} | \{q\}} \
s\chi_{\{\dot
 p\} | \{q\}} (M_1) \ s\chi_{\{\dot p\} | \{q\}} (M_2),
 \label{primera}
 \end{eqnarray}
 in analogy with (\ref{B2}). So, the first way of calculating
 (\ref{primera}) leads to
\begin{eqnarray}
I_{\{p\},\{q\}}(M_1, M_2) =  \left[ \frac{\sigma_{\{p\}}
 \sigma_{\{q \}} }{|p|!  \ |q|!}  \right] ^2 \alpha_{\{\dot p\} |
\{q\}} (str \
 M_1)^{|p|+|q|} (str \ M_2)^{|p|+|q|} + \dots
\label{primerap}
\end{eqnarray}
where  only  the term containing the highest
 power in $(str \ M_1) (str \ M_2)$ has been written.

Now, the second way of
 calculating $I_{\{p\},\{q\}}(M_1, M_2)$ consists in using the expansion
(\ref{term1})
for the
 integrand in (\ref{Ipq})
\begin{eqnarray}
I_{\{p\},\{q\}}(M_1, M_2) =\int [dU]
 \underbrace{s\chi_{\{p\}}((M_1 U M_2
U^\dagger)^\dagger)}_{s\chi_{\{p\}}(M_1 U
 M_2 U^\dagger)} \ s\chi_{\{q\}}(M_1 U M_2 U^\dagger)+ \dots .
\label{seg}
 \end{eqnarray}
and keeping only the highest power term. Next we  combine the
representations
in the RHS
of this relation
\begin{eqnarray}
\int [dU]  \ s\chi_{\{p\}}(M_1 U M_2 U^\dagger) \
s\chi_{\{q\}}(M_1 U M_2 U^\dagger)& \nonumber  \\
= \int [dU] &s\chi_{\{p\} \otimes \{q\}}(M_1
 U M_2 U^\dagger)
\label{segu}
\end{eqnarray}
and subsequently  we use the following  Clebsh-Gordan expansion
arising from
(\ref{ct})
\begin{eqnarray}
s\chi_{\{p\} \otimes \{q\}} (A) = \sum_{\{t\}}^{}{'} \
\rho_{\{t\}}^{\{p\},\{q\}} \
s\chi_{\{t\}}(A).
\label{pxq}
\end{eqnarray}
So we have that
\begin{eqnarray}
\int [dU] \
s\chi_{\{p\}}((M_1 U M_2 U^\dagger) \ s\chi_{\{q\}}(M_1 U M_2
U^\dagger) &
\nonumber \\
 = \sum_{\{t\}}^{}{'}  \  \rho_{\{t\}}^{\{p\},\{q\}} &  \int [dU]
s\chi_{\{t\}}(M_1 U M_2 U^\dagger)
\nonumber
\end{eqnarray}
\begin{eqnarray}
= \sum_{\{t\}}^{}{'} \ \rho_{\{t\}}^{\{p\},\{q\}} \ \alpha_{\{t\}} \
s\chi_{\{t\}}(M_1)
\ s\chi_{\{t\}}(M_2).
\label{segun}
\end{eqnarray}
Therefore, substituting (\ref{segun}) in
(\ref{seg}), we obtain
\begin{eqnarray}
I_{\{p\},\{q\}}(M_1, M_2) = \sum_{\{t\}}^{}{'}
\rho_{\{t\}}^{\{p\},\{q\}} \
\alpha_{\{t\}} \ s\chi_{\{t\}}(M_1) \ s\chi_{\{t\}}(M_2) + \dots ,
\end{eqnarray}
and using
(\ref{term2}) we are left with
\begin{eqnarray}
I_{\{p\},\{q\}}(M_1, M_2)=
\frac{1}{[(|p|+|q|)!]^2}  \sum_{\{t\}}^{}{'}
\rho_{\{t\}}^{\{p\},\{q\}}\alpha_{\{t\}} \
\sigma_{\{t\}}^2
(str \ M_1)^{|p|+|q|} & \nonumber \\
\times (str \ M_2)^{|p|+|q|}  + \dots . &
\label{segunda}
\end{eqnarray}
This is
the result obtained by following the second method of calculation.

Finally we see that comparing the coefficient of the term
$(str \ M_1)^{|p|+|q|}$  \\ $ \times (str \ M_2)^{|p|+|q|}$ of the
two expressions for $I_{\{p\},\{q\}}(M_1, M_2)$, given in
(\ref{primerap}) and
(\ref{segunda}), we obtain the desired result stated in
(\ref{alphamix}) for
the
$\alpha$ coefficients of the mixed representations.

As  a consequence of our results (\ref{alphamix}) and
(\ref{TMENOR}) we  derive
the result
\begin{eqnarray}
\alpha_{\{\dot p\}|\{q\}} = 0, \ \mbox{for $|p|+|q| < mn,$}
\end{eqnarray}
which is  similar to the one in
(\ref{TMENOR}).

Before closing this section we also  observe  that
\begin{eqnarray}
\alpha_{\{\dot p\}|\{q\}} = \alpha_{\{\dot q\}|\{p\}}.
\end{eqnarray}
This property can be obtained from the relation
(\ref{alphamix}) together with the fact that the tensor product of
representations commutes.

\section{Identities for the dimensions of the $U(N)$ representations}
\label{IDU}

The complete procedure already followed for the
determination of the $\alpha_{\{\dot p\} | \{q\}}$ coefficients may be
repeated step by step  for the case of
$U(N)$,   obtaining exactly the same relation (\ref{alphamix}), but  with
the substitution $\alpha_{\{t\}} \rightarrow \frac{1}{d_{\{t\}}}$
everywhere
(see Eq. (\ref{1sobred}))
and also with the replacement $s tr  \rightarrow  tr $. In this way
we obtain
the
remarkable result
\begin{eqnarray}
\frac{1}{d_{\{\dot p\} | \{q\}}} = \left[ \frac{ |p|!  \
|q|!}{(|p|+|q|)!}  \right] ^2 \left[\frac{1}{\sigma_{\{p\}}
\sigma_{\{q \}}}
\right] ^2 \sum_{\{t\}}^{}{'} \
\rho_{\{t\}}^{\{p\},\{q\}} \ \frac{\sigma_{\{t\}}^2}{d_{\{t\}}}.
\label{dimension7}
\end{eqnarray}
for the dimensions of the  irreducible representations of $U(N)$.
In fact,  since  these dimensions  are all  well known from an
independent
calculation
($d_{\{t\}}= \chi_{\{t\}}(I_{N \times N}$)),
Eq.(\ref{dimension7})  provides  an identity  relating the  dimensions of
mixed and undotted representations of this group.
Many examples of the identity (\ref{dimension7}) are shown
explicitly, mutatis
mutandis,   in the  Appendix  \ref{Alfas}.  Let us illustrate this,   for
example, in the
case of  the representation
$\begin{picture}(18,6)(0,0)
\put(0,0){\framebox(6,6){$\cdot$}}
\put(6,0){\framebox(6,6){$\space$}}
\put(12,0){\framebox(6,6){$\space$}}
\end{picture}$ \ .
According to Appendix  \ref{Alfas}  (second row) we have that
\begin{eqnarray}
\frac{1}{d_{\begin{picture}(18,6)(0,0)
\put(0,0){\framebox(6,6){$\cdot$}}
\put(6,0){\framebox(6,6){$\space$}}
\put(12,0){\framebox(6,6){$\space$}}
\end{picture}}} =
\frac{1}{9} \frac{1}{d_{\begin{picture}(18,6)(0,0)
\put(0,0){\framebox(6,6){$\space$}}
\put(6,0){\framebox(6,6){$\space$}}
\put(12,0){\framebox(6,6){$\space$}}
\end{picture}}} + \frac{4}{9} \frac{1}{d_{\begin{picture}(18,6)(0,0)
\put(0,0){\framebox(6,6){$\space$}}
\put(6,0){\framebox(6,6){$\space$}}
\put(0,-6){\framebox(6,6){$\space$}}
\end{picture}}},
\label{exampled}
\end{eqnarray}
where
\begin{eqnarray}
\begin{array}{c}
d_{\begin{picture}(18,6)(0,0)
\put(0,0){\framebox(6,6){$\cdot$}}
\put(6,0){\framebox(6,6){$\space$}}
\put(12,0){\framebox(6,6){$\space$}}
\end{picture}} = \frac{1}{2} N(N+2)(N-1), \quad
d_{\begin{picture}(18,6)(0,0)
\put(0,0){\framebox(6,6){$\space$}}
\put(6,0){\framebox(6,6){$\space$}}
\put(12,0){\framebox(6,6){$\space$}}
\end{picture}} = \frac{1}{6} N(N+1)(N+2),  \nonumber \\
\\
\mbox{} \ \ \ d_{\begin{picture}(18,6)(0,0)
\put(0,0){\framebox(6,6){$\space$}}
\put(6,0){\framebox(6,6){$\space$}}
\put(0,-6){\framebox(6,6){$\space$}}
\end{picture}} = \frac{1}{3}N(N+1)(N-1),
\end{array} \\
\label{formdim}
\end{eqnarray}
according to the  formulas in Table I of  the  Appendix
\ref{charactertable}.
The
reader may
verify that the identity
in (\ref{exampled}) is indeed fulfilled by  expressions  (\ref{formdim}).

\section{Relabeling of the $U(m|n)$ representations}
\label{RUR}

The irreducible representations of $SU(m|n)$ have been characterized
by Bars and Balantekin in terms of the Young tableaux
notation \cite{BARS1}. We referred to this classification in
section 3, when applying it to  the $U(m|n)$
case. In this
notation,  to completely specify each representation,
a set of numbers $(t_1, \ldots, t_k)$  is required, counting  the
number of
boxes in
the corresponding rows of the $\{t\}$ tableau. Contrary to what
happens in the
$U(N)$ ($SU(N)$) case, where the number of rows of the undotted tableau
should not exceed $N$ ($(N-1)$), in the $U(m|n)$ ($SU(m|n)$) case
there is no
restriction for this number  which, in principle,
may be as
large
as wanted  \cite{BARS1}.  So, using the number of boxes on each row as a
labelling
of the $U(m|n)$
representations requires a non definite number of parameters.

Using our formula (\ref{resultcpq}) we will show that it is possible
to choose, at most, $(m+n)$
parameters in order to completly specify the undotted representations
of $U(m|n)$. This is because representations of the type
\begin{center}
\begin{picture}(116,84)(0,0)
\put(0,-16){\framebox(12,12){$\space$}}
\put(12,-16){\framebox(12,12){$\space$}}
\put(0,-28){\framebox(12,12){$\space$}}
\put(12,-28){\framebox(12,12){$\space$}}
\put(0,-40){\framebox(12,12){$\space$}}
\put(0,0){\framebox(12,12){$\space$}}
\put(12,0){\framebox(12,12){$\space$}}
\put(24,0){\framebox(12,12){$\space$}}
\put(0,12){\framebox(12,12){$\space$}}
\put(12,12){\framebox(12,12){$\space$}}
\put(24,12){\framebox(12,12){$\space$}}
\put(40,12){\framebox(12,12){$\space$}}
\put(52,12){\framebox(12,12){$\space$}}
\put(40,0){\framebox(12,12){$\space$}}
\put(52,0){\framebox(12,12){$\space$}}
\put(-109,27){$\{t\}_E =\{mn\}\{p\}=$}
\put(-72,11){$\{q\}^T \{r\}$}
\put(-12,27){$m$}
\put(0,24){\framebox(12,12){$\space$}}
\put(12,24){\framebox(12,12){$\space$}}
\put(24,24){\framebox(12,12){$\space$}}
\put(40,24){\framebox(12,12){$\space$}}
\put(52,24){\framebox(12,12){$\space$}}
\put(0,36){\framebox(12,12){$\space$}}
\put(12,36){\framebox(12,12){$\space$}}
\put(24,36){\framebox(12,12){$\space$}}
\put(40,36){\framebox(12,12){$\space$}}
\put(52,36){\framebox(12,12){$\space$}}
\put(64,36){\framebox(12,12){$\space$}}
\put(0,48){\framebox(12,12){$\space$}}
\put(12,48){\framebox(12,12){$\space$}}
\put(24,48){\framebox(12,12){$\space$}}
\put(40,48){\framebox(12,12){$\space$}}
\put(52,48){\framebox(12,12){$\space$}}
\put(64,48){\framebox(12,12){$\space$}}
\put(76,48){\framebox(12,12){$\space$}}
\put(15,63){$n$}
\put(0,-16){\framebox(12,12){$\space$}}
\put(12,-16){\framebox(12,12){$\space$}}
\put(0,-28){\framebox(12,12){$\space$}}
\put(12,-28){\framebox(12,12){$\space$}}
\put(0,-40){\framebox(12,12){$\space$}}
\put(40,-16){\framebox(12,12){$\space$}}
\put(52,-16){\framebox(12,12){$\space$}}
\put(40,-28){\framebox(12,12){$\space$}}
\put(24,-16){\framebox(12,12){$\space$}}
\put(24,-28){\framebox(12,12){$\space$}}
\end{picture}
\end{center}
\begin{eqnarray}
\label{main-tableau2}
\end{eqnarray}
do not exist, whenever representation $\{r\}$ is  allowed to be placed
there, that is, when
$(\{q\}^T)_i = n$ for every $r_i \neq 0$ and $r_1 \leq p_m$.

To understand this property from our point of view,  let us observe that
formula (\ref{resultcpq}) can be
extended for representations $\{q\}^T \rightarrow \{q\}^T \{r\}$.
In fact,
in Appendix \ref{chi}
we deal with this formula and validate it for
a $\{q\}^T$ tableau having any number of rows and columns, as long as the
Young tableaux rules are kept obeyed. Then for  the
representation $\{t\}_E$ (for which we mean `Extended' $\{t\})$ we
have that
\begin{eqnarray}
s\chi_{\{t\}_E} (M) = (-1)^{|q|+|r|} \ \Sigma (\lambda, \bar{\lambda})
\ \chi_{\{p\}}(\lambda) \ \chi_{\begin{array}{l}
                                   \{q\} \\
                                   \{r\}^T
                             \end{array} } (\bar{\lambda}).
\label{resultcpq3}
\end{eqnarray}
But  any  $U(n)$ tableau having more than $n$ rows is forbidden, i.e.
\begin{eqnarray*}
\chi_{\begin{array}{l}
              \{q\} \\
              \{r\}^T
        \end{array} } (\bar{\lambda}) =0,
\end{eqnarray*}
so that
\begin{eqnarray}
s\chi_{\{t\}_E} (M) = 0.
\label{nullsuper}
\end{eqnarray}
Given that the dimension of a supergroup representation can be
calculated as
$sd_{\{t\}}=s\chi_{\{t\}}(M_0)$, where $M_0$ is given in (\ref{mat-dim}),
 we have
that the dimension for $U(m|n)$ representations of the type $\{t\}_E$ is
0 and therefore they do not exist. This fact was already known in the
literature \cite{BARS-PHYSICA}, but it appears naturally in our  
calculations.

The above observation leads us to propose that
any   legal $U(m|n)$ re\-pre\-sen\-ta\-tion can be completely
characterized
by
\begin{equation}
\{t \Vert s\} \equiv (t_1, \ldots, t_m \Vert s_1, \dots, s_n),
\end{equation}
in such a way  that
\begin{center}
\begin{picture}(72,60)(0,0)
\put(-16.1,0){$t_m$ \framebox(12,12){$\space$}}
\put(12,0){\framebox(24,12){$\cdots$}}
\put(36,0){\framebox(12,12){$\space$}}
\put(52,0){\framebox(12,12){$\space$}}
\put(0,12){\framebox(12,36){$\vdots$}}
\put(12,12){\framebox(24,36){$\vdots$}}
\put(36,12){\framebox(12,36){$\vdots$}}
\put(52,12){\framebox(12,36){$\vdots$}}
\put(64,12){\framebox(12,36){$\vdots$}}
\put(76,36){\framebox(12,12){$\space$}}
\put(36,63){$s_n$}
\put(0,63){$s_1$}
\put(-12.8,48){$t_1$ \framebox(12,12){$\space$}}
\put(12,48){\framebox(24,12){$\cdots$}}
\put(36,48){\framebox(12,12){$\space$}}
\put(52,48){\framebox(12,12){$\space$}}
\put(64,48){\framebox(12,12){$\space$}}
\put(76,48){\framebox(12,12){$\space$}}
\put(88,48){\framebox(12,12){$\space$}}
\put(0,-16){\framebox(12,12){$\space$}}
\put(12,-16){\framebox(24,12){$\cdots$}}
\put(36,-16){\framebox(12,12){$\space$}}
\put(0,-28){\framebox(12,12){$\space$}}
\put(12,-28){\framebox(24,12){$\cdots$}}
\put(0,-40){\framebox(12,12){$\space$}}
\end{picture}
\begin{eqnarray}
\label{main-tableau21}
\end{eqnarray}
\end{center}
\vspace{0.8cm}
\noindent where $(t_1, \dots, t_m)$ is a $U(m)$ tableau denoting
the number
of boxes in
the first $m$ rows of $\{t \Vert s\}$, while $(s_1, \ldots, s_n)$
is a $U(n)$
tableau denoting the number of boxes of the first $n$ columns of
 $\{t \Vert s\}$. These set of numbers completely specifies the existing
 undotted representations of $U(m|n)$.

If the $t_i$'s and the $s_j$'s in (\ref{main-tableau21}) satisfy
respectively
\begin{eqnarray}
t_i \geq n, \ s_j \geq m, \ (i=1, \ldots, m; \ j=1, \ldots, n),
\label{acond}
\end{eqnarray}
then these numbers are completly independent. In
this case $\{t \Vert s\}$ is a tableau of the type $\{\tilde{t}\}$
 in (\ref{main-tableau}). But if the $t_i$'s and the $s_j$'s do not  
all obey
(\ref{acond}) then they will not be all independent. In fact, if
$\{t \Vert s\}$ is such that every box of the tableaux
is contained in the $\{mn\}$
tableau,
then knowing all the $t_i$'s is completely equivalent to knowing all the
$s_j$'s. Anyway, it is still true that knowing the $m+n$ numbers
$(t_1, \dots, t_m)$ and $(s_1, \ldots, s_n)$ (assumed to be given
unambiguously and consistently) is enough to specify
any $U(m|n)$ undotted representation.


Now, the analogue  happens when considering purely dotted
representations.
Equation (\ref{resultcpq}) is also valid for dotted representations since
the corresponding derivation can be
completely repeated for this case (the character and
supercharacter expansions of the ordinary and supersymmetric HCIZ
integral may
be directly obtained for purely dotted representations).
So following exactly the same
arguments we are led to state that every $U(m|n)$ dotted
representation can be
completely specified by the notation
$\{\dot{t} \Vert \dot{s} \} \equiv
(\dot{t}_1, \dots, \dot{t}_m \Vert \dot{s}_1, \dots, \dot{s}_n)$.
The pictorical
tableau would be the same as in (\ref{main-tableau21}) but with all
boxes dotted.

In the case of mixed representations, the undotted and the dotted
parts will
follow separately the previously established rules and
the tableau will be abbreviated as
$\{ \dot{p} \Vert \dot{q} \}|\{ u \Vert v \}$.

\vfill
\eject

\section{Appendices}


\subsection{The supersymmetric HCIZ integral }
\label{SUSYHCIZ}

The basic tool we have used to determine the integration
properties of the
supergroup
$U(m|n)$  in this work is the supersymmetric extension of the
Harish-Chandra-Itzykson-Zuber ( SUSY HCIZ) integral defined by 
\cite{Nosotros1, GUHR2}
\begin{eqnarray}
\tilde{I}(M_1,M_2;\beta)= \int [dU] e^{\beta str(M_1 U M_2 U^\dagger)},
\label{SUSYIZUBER}
\end{eqnarray}
where $M_1$ and $M_2$ are hermitian $(m+n)
\times (m+n)$ supermatrices, the integration is carried over the
supergroup
$U(m|n)$ and `str' means the supertrace operation.  This extension is
made in
complete analogy with the ordinary HCIZ integral which is \cite{IZ}
\begin{eqnarray}
I(N_1, N_2;\beta) = \int [dU] e^{\beta tr(N_1 U N_2 U^\dagger)},
\label{ordinary}
\end{eqnarray}
where $N_1$ and $N_2$ are $N \times
N$ hermitian matrices, and the integration is carried over the
group $U(N)$.

The calculation of the SUSY HCIZ integral has been made by
following analogous
steps to those taken by Itzykson and Zuber in the ordinary $U(N)$ case
\cite{Nosotros1, GUHR2}.  In this approach, the integral is not
calculated
directly, but it is found as the solution of a differential
equation.  In the
ordinary case this procedure is known as `the difussion equation
method', but
in
our case it was transformed to `the Schr\"odinger equation method',
in which,
for convergence reasons we incorporated an imaginary factor `i' to the
difussion
equation \cite{Nosotros1}.

The result for the calculation of the SUSY HCIZ integral is
\begin{eqnarray}
\tilde{I}(M_1, M_2; \beta) = \Sigma (\lambda_1,{\bar\lambda}_1) \Sigma
(\lambda_2,{\bar\lambda}_2) \ \beta^{mn} \times (\beta)^ {- \frac
{m(m-1)}{2}}
(- \beta)^{- \frac{n(n-1)}{2}} \times \nonumber \\ \times
\prod_{p=1}^{m-1} p!
\prod_{q=1}^{n-1} q!  \frac {det(e^{ \beta \lambda_{1i} \lambda_{2j}
})}{\Delta
(\lambda_1) \Delta (\lambda_2)} \frac {det(e^{-\beta
{\bar\lambda}_{1\alpha}
{\bar\lambda}_{2\beta}} )} {\Delta ({\bar\lambda}_1) \Delta
({\bar\lambda}_2)},
\label{expresion} \end{eqnarray}
where the diagonal supermatrices
$\Lambda_i$
($i=1,2$) contain the eigenvalues of the respective $(m+n)\times(m+n)$
 hermitian supermatrices $M_i$ ($i=1,2$) (see (\ref{diagonalsup}) for the
 conventions).

Here, $\Delta$ is the usual Vandermonde determinant
\begin{eqnarray} \Delta
(\lambda) = \prod_{i>j} (\lambda_i - \lambda_j) ,\ \ \Delta
(\bar\lambda) =
\prod_{ \alpha> \beta} ({\bar\lambda}_{\alpha} - {\bar\lambda}_{\beta})
\label{Delta}
\end{eqnarray}
\noindent and the new function that appears is
\begin{eqnarray*}
\Sigma (\lambda,\bar\lambda) =
\prod_{i=1}^m \prod_{\alpha=1}^n
( \lambda_i - {\bar\lambda}_{\alpha} ).
\end{eqnarray*}
We observe that the polynomial $\Sigma (\lambda,\bar\lambda)$ is
completely
symmetric
under
independent permutations of the $\lambda_i$'s and the
$\bar{\lambda}_\alpha$'s.

The expression (\ref{expresion}) is completely determined up to a
normalization
factor related to that of the measure $[dU]$ of the supergroup.
This situation
is analogous to the standard IZ case where the required factor can
be fixed
directly from the corresponding expression by taking the limit
$\Lambda_1,
\Lambda_2 \rightarrow 0$ in a convenient way and demanding $\int
[dU] = 1$, for
example.  This procedure leads to the correct factors in Eq.(3.4) of
Ref.\cite{IZ}.  In our case, a similar limiting procedure leads to the
conclusion that $\int [dU] \equiv 0,$ precisely due to the
appearance of the
$\Sigma (\lambda, \bar{\lambda})$ functions in the numerator.  This
is not an
unexpected result since we are dealing with odd Grassmann numbers.
For this
reason we have chosen the normalization factor in such a way that
\begin{eqnarray}
\tilde{I} (
M_1, M_2;\beta) = \Sigma (\lambda_1, {\bar\lambda}_1) \Sigma
(\lambda_2,{\bar\lambda}_2) \beta^{mn} I(\lambda_1,\lambda_2;\beta)
I({\bar\lambda}_1,{\bar\lambda}_2;-\beta) ,
\label{eq-import}
\end{eqnarray}
where
\begin{eqnarray*}
\tilde{I}(\Lambda_1, \Lambda_2;\beta):  & \mbox{HCIZ integral over
$U(m|n)$} \\
I(\lambda_1,\lambda_2;\beta):  & \mbox{HCIZ integral over $U(m)$} \\
I(\bar{\lambda}_1,\bar{\lambda}_2;-\beta):  & \mbox{HCIZ integral
over $U(n),$}
\end{eqnarray*}
and the HCIZ integral in (\ref{ordinary}) is given by
\begin{equation}
I(N_1,N_2;\beta) = \int [dU] e^{\beta tr(N_1 U N_2 U^\dagger)} =
\beta^{-N(N-1)/2}
\prod_{p=1}^{N-1} p!  \frac {det(e^{\beta \lambda_{1i}
\lambda_{2,j}})} {\Delta
(\lambda_1) \Delta (\lambda_2)}.
\end{equation}

We comment that the derivation of the expression
(\ref{expresion}) has been performed for purely imaginary $\beta$
in order to
guarantee the convergence of the method.  Since both sides of
Eq.(\ref{expresion}) exist for every complex $\beta$ we have made an
analytic
continuation of the result to all the $\beta$ complex plane.

\subsection{Expression for $s\chi_{\{{\tilde t}\}}(M)$}
\label{chi}

Here we show that  the supercharacter of the particular representation
\begin{eqnarray}
\{\tilde{t}\} =\begin{array}{l}
                   \{mn\} \{p\} \\
                   \{q\}^T
             \end{array} ,
\label{TABLEAU}
\end{eqnarray}
which is pictorically shown in (\ref{main-tableau}), has the
compact expression
\begin{eqnarray}
s\chi_{\{{\tilde t}\}} (M) = (-1)^{|q|} \Sigma (\lambda, \bar{\lambda})
\chi_{\{p\}}(\lambda) \chi_{\{q\}} (\bar{\lambda}).
\label{resultcpq1}
\end{eqnarray}
This formula was previously
stated in
Ref.\cite{Nosotros2} and we now present the complete proof of it.

The basic idea of the proof is to   start from  Eq.(\ref{schip})
\begin{eqnarray}
s\chi_{\{mn\}\{p\}}(M)=\Sigma(\lambda,\bar{\lambda})
\chi_{\{p\}}(\lambda),
\label{start}
\end{eqnarray}
valid for every representation $\{p\} \in U(m)$, and
subsequently  to perform an  induction process
in the number of boxes of the representation  $\{q\} \in U(n)$.
We will work with the simplified notation (\ref{TABLEAU}) instead of the
one in (\ref{main-tableau}).

Let $\{q\}^T=\{v\}=(v_1, \ldots, v_a)$ be the tableau which is
placed in the
bottom left of $\{mn\}\{p\}$ in (\ref{TABLEAU}). Our proof will go in two
steps. The first one consists in making induction in the number of
boxes of
the last row of $\{v\}$, that is, $v_a$. The second step consists in
assuming (\ref{resultcpq1}) to be valid
for $\{v\}$ and showing that is also valid for $\{v\}'$, which is
constructed
from $\{v\}$ by  adding an extra row consisting in only one box, that is,
$\{v\}'=(v_1, \ldots, v_a, 1)$. Thus, both proofs imply  that the
$\{q\}^T=\{v\}$ tableau may be as wide and as long as the Young
tableaux rules
allow.

\noindent ${\bf (i)}$ Here we perform the induction
process  in the number of boxes of the last row of
$\{v_a\}$. We assume that (\ref{resultcpq1})
is valid for a tableau $\{q\}^T=\{v\}$, with
\begin{eqnarray}
v_i\leq n \ (i=1, \ldots, a-1) \nonumber \\
\mbox{and for $v_a=0, \ldots, V_a < v_{a-1}.$}
\label{ra}
\end{eqnarray}
Our task consists in showing that it is also valid for
$v_a=V_a+1$.
Let $\{v_o\} = (v_1,\ldots, v_{a-1})$.
We start  by multiplying
\begin{eqnarray}
s\chi_{       \begin{array}{l}
              \{mn\}\{p\} \\ \{v_0\} \ \ \ \ \
              \end{array} } (M) = (-1)^{|v_0|} \Sigma (\lambda,
\bar{\lambda})
\chi_{\{p\}}(\lambda) \chi_{\{v_0\}^T} (\bar{\lambda}),
\end{eqnarray}
by the expression \cite{BARS2},
\begin{eqnarray}
s\chi_{(V_a+1)}(M) = \sum_{k=0}^{V_a+1} (-1)^k \chi_{(V_a+1-k)}(\lambda)
\chi_{(k)^T }(\bar{\lambda}),
\label{spec-for}
\end{eqnarray}
where $(s)$ and $(s)^T$ denote the completely symmetric and the
completely
antisymmetric tableau, respectively, both with s boxes. Using the Young
tableaux
rules for multiplying representations we have
\begin{eqnarray}
s\chi_{  \left(     \begin{array}{l}
              \{mn\}\{p\} \\ \{v_0\} \ \ \ \ \
              \end{array} \right) \otimes (V_a+1)} (M) = (-1)^{|v_0|}
\Sigma
(\lambda, \bar{\lambda}) \times \nonumber \\
\times \sum_{k=0}^{V_a+1} (-1)^k
\chi_{\{p\} \otimes (V_a+1-k)}(\lambda) \chi_{\{v_0\}^T \otimes (k)^T}
(\bar{\lambda}) \\
\Rightarrow \sum_{k=0}^{V_a+1} s\chi_{ \begin{array}{l}
              \{mn\} (\{p\} \otimes (V_a+1-k)) \\ (\{v_0\} \otimes
(k)) \ \ \ \ \
              \end{array} } (M) = (-1)^{|v_0|} \Sigma (\lambda,
\bar{\lambda})
\times \nonumber \\
\times \sum_{k=0}^{V_a+1} (-1)^k
\chi_{\{p\} \otimes (V_a+1-k)}(\lambda) \chi_{\{v_0\}^T \otimes (k)^T}
(\bar{\lambda}).
\label{mientras}
\end{eqnarray}
Now, we separate the $(V_a+1)$th  term in both sides
\begin{eqnarray}
\sum_{k=0}^{V_a} s\chi_{   \begin{array}{l}
              \{mn\} (\{p\} \otimes (V_a+1-k)) \\ (\{v_0\} \otimes
(k))
              \end{array}  } (M) + s\chi_{   \begin{array}{l}
              \{mn\} \{p\} \\ (\{v_0\} \otimes (V_a+1))
              \end{array} } (M)  = \nonumber \\
 (-1)^{|v_0|} \Sigma (\lambda, \bar{\lambda}) \sum_{k=0}^{V_a} (-1)^k
\chi_{\{p\} \otimes (V_a+1-k)}(\lambda) \chi_{\{v_0\}^T \otimes (k)^T}
(\bar{\lambda})&  \nonumber \\
+ (-1)^{|v_0|+V_a+1} \Sigma (\lambda, \bar{\lambda})
\chi_{\{p\}}(\lambda)
\chi_{\{v_0\}^T \otimes (V_a+1)^T} (\bar{\lambda}).& \nonumber \\
\label{last}
\end{eqnarray}
Using the property $(\{a\} \otimes \{b\})^T = \{a\}^T \otimes \{b\}^T$
and the fact that the representations $(\{v_0\} \otimes (k))$ are all
of the type $\{v\}$ (for which the hypothesis of induction (\ref{ra})
 is valid), the sums in both sides of (\ref{last}) are cancelled,
leading to
\begin{eqnarray}
s\chi_{       \begin{array}{l}
              \{mn\} (\{p\}) \\
               (\{v_0\} \otimes (V_a+1)) \
              \end{array} } (M) =& (-1)^{|v_0|+V_a+1} \
\Sigma (\lambda, \bar{\lambda}) \times \nonumber \\
& \times \chi_{\{p\}}(\lambda)
  \chi_{(\{v_0\} \otimes (V_a+1))^T} (\bar{\lambda}).
\label{casi1}
\end{eqnarray}
Using the Young tableaux rules we have that
\begin{eqnarray}
\{v_0\} \otimes (V_a+1) = {       \begin{array}{l}
                                  \{v_0\}  \\ (V_a+1)
                                  \end{array} }
\oplus {\sum_{l_1, \ldots, l_a} } '
(v_1+l_1, \ldots,
v_{a-1}+l_{a-1}, l_a),
\label{product2}
\end{eqnarray}
where ${ \begin{array}{l}
         \{v_0\}  \\ (V_a+1)
         \end{array}}\equiv (v_1, \ldots, v_{a-1}, V_a +1)$
and the prime means that the sum is restricted to $i) \
l_1+\ldots+l_a=V_a+1$,
where all the $l_i$'s are non negative integers; $ii) \ l_a \leq V_a$ and
$iii) \ v_{i} \geq v_{i+1}+l_{i+1}, \mbox{for} \ i=1, \ldots a-1.$

Therefore, using (\ref{product2}) in (\ref{casi1}) we have that
\begin{eqnarray}
\sum_{l_1, \ldots, l_a}{}^{'' } s\chi_{ \begin{array}{l}
              \{mn\} \{p\} \\ (v_1+l_1, \ldots,
v_{a-1}+l_{a-1},l_a)
              \end{array} }(M) + s\chi_{ \begin{array}{l}
              \{mn\} \{p\} \\ \{v_0\} \\ (V_a+1)
              \end{array} }(M) = \nonumber \\
\sum_{l_1, \ldots, l_a}{}^{''} (-1)^{|v_0|+V_a+1} \Sigma (\lambda,
\bar{\lambda})
\chi_{\{p\}}(\lambda) \chi_{(v_1+l_1, \ldots, v_{a-1}+l_{a-1}, l_a)^T}
(\bar{\lambda})& \nonumber \\
+ (-1)^{|v_0|+V_a+1} \Sigma (\lambda, \bar{\lambda})
\chi_{\{p\}}(\lambda) \chi_{\left(  \begin{array}{l}
         \{v_0\}  \\ (V_a+1)
         \end{array}   \right)^T} (\bar{\lambda})&. \nonumber \\
\label{casi2}
\end{eqnarray}
Here the double prime means that the summation is further restricted to
$l_a < V_a+1$. In virtue of the hipothesis of induction the sums
that appear in
both
sides of this equation are equal and we are left with
\begin{eqnarray}
s\chi_{ \begin{array}{l}
              \{mn\} \{p\} \\
              \left( \begin{array}{l}
                    \{v_0\} \\
                    (V_a+1)
                    \end{array}
              \right)
              \end{array}}(M) = (-1)^{|v_0|+V_a+1} \Sigma (\lambda,
\bar{\lambda})\chi_{\{p\}} (\lambda) \chi_{\left( \begin{array}{l}
         \{v_0\}  \\ (V_a+1)
         \end{array}\right)^T} (\bar{\lambda}),\nonumber \\
\label{formula2}
\end{eqnarray}
which ends this part of the proof.

\noindent ${\bf (ii)}$ We will now prove that if (\ref{resultcpq1})
is valid for $v_i \leq n$ ($i=1, \ldots a$), then it is also true that
\begin{eqnarray}
s\chi_{       \begin{array}{l}
              \{mn\} \{p\} \\ \{v\} \\ \Box
              \end{array} } (M) = (-1)^{|v|+1} \Sigma (\lambda,
\bar{\lambda})
\chi_{\{p\}} (\lambda) \chi_{\left( \begin{array}{l}
         \{r\}  \\ \Box
         \end{array}   \right)^T} (\bar{\lambda}).
\label{ind2}
\end{eqnarray}
We will follow  very similar steps  to those  in $(\bf i)$. We multiply
(\ref{resultcpq1}) by  $s\chi_{\Box} (M) = \chi_{\Box} (\lambda)-
\chi_{\Box}(\bar{\lambda})$, obtaining
\begin{eqnarray}
s\chi_{  \left(     \begin{array}{l}
              \{mn\}\{p\} \\ \{v\} \ \ \ \ \
              \end{array} \right) \otimes \Box}(M) = (-1)^{|v|} \Sigma
(\lambda, \bar{\lambda})
\chi_{\{p\} \otimes \Box }(\lambda) \chi_{\{v\}^T} (\bar{\lambda}) +
\nonumber
\\
+ (-1)^{|v|+1} \Sigma (\lambda, \bar{\lambda})
\chi_{\{p\} }(\lambda) \chi_{\{v\}^T \otimes \Box} (\bar{\lambda})
\end{eqnarray}
\begin{eqnarray}
\Rightarrow s\chi_{  \begin{array}{l}
              \{mn\} (\{p\}  \otimes \Box) \\ \{v\}
              \end{array}}(M) + s\chi_{  \begin{array}{l}
              \{mn\} \{p\} \\ ( \{v\} \otimes \Box)
              \end{array}}(M)=    \nonumber \\
 (-1)^{|v|} \Sigma (\lambda, \bar{\lambda})
\chi_{\{p\} \otimes \Box }(\lambda) \chi_{\{v\}^T} (\bar{\lambda}) +
(-1)^{|v|+1} \Sigma (\lambda, \bar{\lambda})
\chi_{\{p\} }(\lambda) \chi_{(\{v\} \otimes \Box)^T} (\bar{\lambda})
\nonumber
\\
\end{eqnarray}
Considering (\ref{resultcpq1}) for the case
$\{p\} \rightarrow \{p\} \otimes \Box$, the first term in both sides is
the same and after cancelling it we have
\begin{eqnarray}
s\chi_{  \begin{array}{l}
              \{mn\} \{p\} \\ ( \{v\} \otimes \Box)
              \end{array}}(M)=
(-1)^{|v|+1} \Sigma (\lambda, \bar{\lambda})
\chi_{\{p\} }(\lambda) \chi_{(\{v\} \otimes \Box)^T} (\bar{\lambda}).
\label{casi3}
\end{eqnarray}
Next we use the analogue formula to
(\ref{product2}) which  is
\begin{eqnarray}
\{v\} \otimes \Box = {       \begin{array}{l}
              \{v\}  \\ \Box
              \end{array} }  \oplus \sum_{j_1, \ldots, j_a} {}^{'''}
(v_1+j_1, \ldots,
v_a+j_a),
\label{product3}
\end{eqnarray}
where the triple prime indicates  the restrictions that  the
$j_i$'s are non
negative integers satisfying
$j_1 + \ldots + j_a = 1$ together with
$v_{i} \geq v_{i+1} + j_{i+1}, \ (i=1, \ldots ,a-1)$.
Using (\ref{product3}) in (\ref{casi3}) we have that
\begin{eqnarray}
\sum_{j_1, \ldots, j_a}{}^{''' }s\chi_{ \begin{array}{l}
              \{mn\} \{p\} \\ (v_1+j_1, \ldots, v_a+j_a)
              \end{array}}(M) + s\chi_{ \begin{array}{l}
              \{mn\} \{p\} \\ \{v\} \\ \Box
              \end{array}}(M) = \nonumber \\
\sum_{j_1, \ldots, j_a}{}^{''' } (-1)^{|v|+1} \Sigma (\lambda,
\bar{\lambda})
\chi_{\{p\}}(\lambda) \chi_{(v_1+j_1, \ldots, v_a+j_a)^T}
(\bar{\lambda}) +
\nonumber \\
+ (-1)^{|v|+1} \Sigma (\lambda, \bar{\lambda})
\chi_{\{p\}}(\lambda) \chi_{\left(  \begin{array}{l}
         \{v\}  \\ \Box
         \end{array}   \right)^T} (\bar{\lambda}).
\label{casi4}
\end{eqnarray}
In virtue of the hipothesis of induction
the sums in both sides are the same and we are left with the desired
result.

\subsection{Supercharacter of mixed representations}
\label{smixrep}


The general expression for the supercharacter of a mixed representation
of the supergroup $GL(m|n)$ is the complicated expression given by
\begin{eqnarray}
s\chi_{\{\dot a \}| \{ b \}} (A) = \sum_{l=0}^{k_{\{a\},\{b\}}}
{\sum_{\{m\}}}''
{\sum_{\{n\}}}''  \delta (r_1 + 2 r_2 + \ldots + l r_l - l)&
\nonumber \\
\times \sum_{r_1, \ldots, r_l} \
\phi_{l,\{m\},\{n\},\{r\}}^{\{a\},\{b\}} \
\prod_{i=1}^{l} [(str(A A^\dagger)^i]^{r_i} \ s\chi_{\{\dot m\}} (A)
s\chi_{\{n\}} (A), &
\label{expresion1}
\end{eqnarray}
where $A$ is an arbitrary $(m+n) \times (m+n)$ supermatrix,
$|m|=|a|-l$,
$|n|=|b|-l$ and $k_{\{a\},\{b\}}= min\{|a|,|b|\}$.
The coefficients  $\phi_{l,\{m\},\{n\},\{r\}}^{\{a\},\{b\}}$ are
known for all
representations $\{a\}$ and $\{b\}$ of $GL(m)$ and $GL(n)$,
respectively. Again,
the double prime on each summation is to remind the reader of the
constraints
over
which the sumations are performed.

In particular
\begin{eqnarray}
\phi_{0,\{a\},\{b\},\{0\}}^{\{a\},\{b\}} = 1,
\end{eqnarray}
which corresponds to the terms in (\ref{expresion1}) which do not contain
any factor $[str \ (A A^\dagger)^j]^{r_j}$. This term is precisely
the one
that we consider in Eq.(\ref{term1}).

The formula
(\ref{expresion1})
is a generalization of the expression appearing in Ref. \cite{BARS0},
which correspond to
the superunitary case where $A A^\dagger=1$

Simple examples of the formula (\ref{expresion1}) are
\begin{eqnarray}
s\chi_{\begin{picture}(12,6)(0,0)
\put(0,0){\framebox(6,6){$\cdot$}}
\put(6,0){\framebox(6,6){$\space$}}
\end{picture}} (A) = s\chi_{\begin{picture}(6,6)(0,0)
\put(0,0){\framebox(6,6){$\cdot$}}
\end{picture}}(A) s\chi_{\begin{picture}(6,6)(0,0)
\put(0,0){\framebox(6,6){$\space$}}
\end{picture}}(A)   - \frac{1}{m-n} str(A A^\dagger),
\label{ex-1}
\end{eqnarray}
\begin{eqnarray}
s\chi_{\begin{picture}(18,6)(0,0)
\put(0,0){\framebox(6,6){$\cdot$}}
\put(6,0){\framebox(6,6){$\space$}}
\put(12,0){\framebox(6,6){$\space$}}
\end{picture}} (A) = s\chi_{\begin{picture}(6,6)(0,0)
\put(0,0){\framebox(6,6){$\cdot$}}
\end{picture}} (A) s\chi_{\begin{picture}(12,6)(0,0)
\put(0,0){\framebox(6,6){$\space$}}
\put(6,0){\framebox(6,6){$\space$}}
\end{picture}} (A) -\frac{1}{m-n} str(AA^\dagger)
s\chi_{\begin{picture}(6,6)(0,0)
\put(0,0){\framebox(6,6){$\space$}}
\end{picture}} (A),  \\
s\chi_{\begin{picture}(12,16)(0,0)
\put(0,6){\framebox(6,6){$\cdot$}}
\put(6,6){\framebox(6,6){$\space$}}
\put(6,0){\framebox(6,6){$\space$}}
\end{picture}} (A) = s\chi_{\begin{picture}(6,6)(0,0)
\put(0,0){\framebox(6,6){$\cdot$}}
\end{picture}} (A) s\chi_{\begin{picture}(6,16)(0,0)
\put(0,0){\framebox(6,6){$\space$}}
\put(0,6){\framebox(6,6){$\space$}}
\end{picture}} (A) - \frac{1}{m-n} str(AA^\dagger)
s\chi_{\begin{picture}(6,6)(0,0)
\put(0,0){\framebox(6,6){$\space$}}
\end{picture}} (A). \nonumber \\
\end{eqnarray}

The reader may verify that these expressions coincide with the ones
of Ref. \cite{BARS0} when $A$ is a unitary supermatrix.

We are not going to perfom here the derivation of (\ref{expresion1}).
Instead,
we will present a simple example which illuminates the general
procedure. Let us take  the case
$\{\dot a\}|\{b\} = \begin{picture}(12,6)(0,0)
\put(0,0){\framebox(6,6){$\cdot$}}
\put(6,0){\framebox(6,6){$\space$}}
\end{picture} \  . $
In order to construct the result for $s\chi_{\begin{picture}(12,6)(0,0)
\put(0,0){\framebox(6,6){$\cdot$}}
\put(6,0){\framebox(6,6){$\space$}}
\end{picture}} (A)$ given in  (\ref{ex-1}) we  consider
\begin{eqnarray}
{\cal D}_{ac,bd}^{\begin{picture}(12,6)(0,0)
\put(0,0){\framebox(6,6){$\cdot$}}
\put(6,0){\framebox(6,6){$\space$}}
\end{picture} } (A) = {\cal D} _{ac,bd}^{\begin{picture}(6,6)(0,0)
\put(0,0){\framebox(6,6){$\cdot$}}
\end{picture}  \times \begin{picture}(6,6)(0,0)
\put(0,0){\framebox(6,6){$\space$}}
\end{picture}} (A) - \frac{1}{m-n} \delta_{bd} (-1)^{\epsilon_e} {\cal
D}_{ac,ee}^{\begin{picture}(6,6)(0,0)
\put(0,0){\framebox(6,6){$\cdot$}}
\end{picture}  \times \begin{picture}(6,6)(0,0)
\put(0,0){\framebox(6,6){$\space$}}
\end{picture}} (A) .
\label{one-one}
\end{eqnarray}
The above expression is obtained  starting from the fundamental
representations
\begin{eqnarray}
{\cal D} _{ij}^{\begin{picture}(6,6)(0,0)
\put(0,0){\framebox(6,6){$\space$}}
\end{picture}} (A) = A_{ij} \qquad {\cal D}
_{ij}^{\begin{picture}(6,6)(0,0)
\put(0,0){\framebox(6,6){$\cdot$}}
\end{picture}} (A) = (-1)^{\epsilon_i (\epsilon_i + \epsilon_j)}
A_{ij}^*,
\end{eqnarray}
and imposing the representation
$\begin{picture}(12,6)(0,0)
\put(0,0){\framebox(6,6){$\cdot$}}
\put(6,0){\framebox(6,6){$\space$}}
\end{picture}$ to be irreducible. The representation
$\begin{picture}(6,6)(0,0)
\put(0,0){\framebox(6,6){$\cdot$}}
\end{picture}  \times \begin{picture}(6,6)(0,0)
\put(0,0){\framebox(6,6){$\space$}}
\end{picture}$
is given by
\begin{eqnarray}
{\cal D} _{ac,bd}^{\begin{picture}(6,6)(0,0)
\put(0,0){\framebox(6,6){$\cdot$}}
\end{picture}  \times \begin{picture}(6,6)(0,0)
\put(0,0){\framebox(6,6){$\space$}}
\end{picture} } (A) = (-1)^{(\epsilon_a + \epsilon_c)(\epsilon_a +
\epsilon_b)}
A_{ba}^\dagger A_{cd},
\end{eqnarray}
according to the general rule described in section \ref{BPSR}.

The construction of (\ref{one-one}) leads to
\begin{eqnarray}
{\cal D}_{ac,bd}^{\begin{picture}(12,6)(0,0)
\put(0,0){\framebox(6,6){$\cdot$}}
\put(6,0){\framebox(6,6){$\space$}}
\end{picture} } (A) = (-1)^{\epsilon_a + \epsilon_c)(\epsilon_a +
\epsilon_b)}
A_{ba}^\dagger A_{cd}  - \frac{1}{m-n} \delta_{bd} (-1)^{\epsilon_e}(A
A^\dagger)_{ca}.
\label{one-one-two}
\end{eqnarray}
Calculating the supercharacter
$s\chi_{\begin{picture}(12,6)(0,0)
\put(0,0){\framebox(6,6){$\cdot$}}
\put(6,0){\framebox(6,6){$\space$}}
\end{picture} } (A) = \sum (-1)^{\epsilon_a + \epsilon_c} {\cal D}
_{ac,ac}^{\begin{picture}(12,6)(0,0)
\put(0,0){\framebox(6,6){$\cdot$}}
\put(6,0){\framebox(6,6){$\space$}}
\end{picture} } (A)$,we obtain (\ref{ex-1}).

Let us observe that in  order to get  (\ref{ex-1}) we have begun from
(\ref{one-one}), which is the product of the representations
$\begin{picture}(6,6)(0,0)
\put(0,0){\framebox(6,6){$\cdot$}}
\end{picture}$ and $\begin{picture}(6,6)(0,0)
\put(0,0){\framebox(6,6){$\space$}}
\end{picture}$
to which we have substracted a similar term with a repeated index
$e$. This
index contraction
produces  the term $A A^\dagger$ in
(\ref{one-one-two}) and subsequently it  becomes $str (A A^\dagger)$,
after calculating the supercharacter.

When the same procedure is applied to more complicated cases like
that of the representation
$\begin{picture}(30,16)(0,0)
\put(0,0){\framebox(6,6){$\cdot$}}
\put(6,0){\framebox(6,6){$\cdot$}}
\put(12,0){\framebox(6,6){$\cdot$}}
\put(18,0){\framebox(6,6){$\space$}}
\put(24,0){\framebox(6,6){$\space$}}
\end{picture}$,
we will obtain an expression of the type
\begin{eqnarray}
s\chi_{\begin{picture}(24,16)(0,0)
\put(0,0){\framebox(6,6){$\cdot$}}
\put(6,0){\framebox(6,6){$\cdot$}}
\put(12,0){\framebox(6,6){$\cdot$}}
\put(18,0){\framebox(6,6){$\space$}}
\put(24,0){\framebox(6,6){$\space$}}
\end{picture}} (A) = s\chi_{\begin{picture}(18,10)(0,0)
\put(0,0){\framebox(6,6){$\cdot$}}
\put(6,0){\framebox(6,6){$\cdot$}}
\put(12,0){\framebox(6,6){$\cdot$}}
\end{picture}} (A)
s\chi_{\begin{picture}(6,16)(0,0)
\put(0,0){\framebox(6,6){$\space$}}
\put(6,0){\framebox(6,6){$\space$}}
\end{picture}} (A) +
a \ str(AA^\dagger) s\chi_{\begin{picture}(12,6)(0,0)
\put(0,0){\framebox(6,6){$\cdot$}}
\put(6,0){\framebox(6,6){$\cdot$}}
\end{picture}} (A)
s\chi_{\begin{picture}(6,6)(0,0)
\put(0,0){\framebox(6,6){$\space$}}
\end{picture}} (A) + \nonumber \\
+ (b [str(A A^\dagger)]^2 + c \ str (A A^\dagger)^2 ) \
s\chi_{\begin{picture}(6,6)(0,0)
\put(0,0){\framebox(6,6){$\cdot$}}
\end{picture}} (A),
\end{eqnarray}
where the coefficients  $a$, $b$ y $c$ take known numerical values.
When the same procedure is extended to the general case, one obtains
the formula (\ref{expresion1}).

\subsection{Character and Supercharacter tables of
$GL(N)$ and $GL(m|n)$}
\label{charactertable}

The character of any  $U(N)$ representation may be written in terms
of traces
of powers of the fundamental ordinary and complex representations,
$U$ and $\bar{U}$.

A general formula for the character of an undotted representation $\{t\}$
is \cite{Littlewood}
\begin{eqnarray}
\chi_{\{t\}}(U) = \frac{1}{|t|!} \sum_{a_1, \ldots, a_{|t|}=0}^{|t|}
\delta (a_1 + 2 a_2 + \ldots + |t| a_{|t|} - |t|) h_{(a)}
\chi_{(a)}^{\{t\}} \prod_{i=1}^{|t|} (trU^i)^{a_i},\nonumber \\
\label{car-formula}
\end{eqnarray}
where the $\chi_{(a)}^{\{t\}}$ coefficients are the characters
of the symmetric group of degree $|t|$, $S_{|t|}$, and
\begin{eqnarray}
h_{(a)} = \frac{|t|!}{1^{a_1} a_1! \ 2^{a_2} a_2! \ldots |t|^{a_{|t|}}
a_{|t|}!}
\end{eqnarray}
is the order of class $\{a\}$ of $S_{|t|}$.

The character of a dotted representation has exactly the same expression
(\ref{car-formula}), but replacing $U \rightarrow U^\dagger$.

For the character of a mixed representation  see  Ref. \cite{BARS0}
and  our  Appendix  \ref{smixrep}, replacing
 supertrace for  trace, whenever it is necessary.
Table I is constructed with these ingredients.

Replacing trace by supertrace \cite{BARS2},
we obtain the analogue of formula (\ref{car-formula}) for the
supercharacter of  the representation ${\{t\}}$ of $GL(m|n)$:
\begin{eqnarray}
s\chi_{\{t\}}(U) = \frac{1}{|t|!} \sum_{a_1, \ldots, a_{|t|}=0}^{|t|}
\delta (a_1 + 2 a_2 + \ldots + |t| a_{|t|} - |t|) h_{(a)}
\chi_{(a)}^{\{t\}} \prod_{i=1}^{|t|} (strU^i)^{a_i}.\nonumber \\
\label{supercar-formula}
\end{eqnarray}
Some examples of this formula appear in Table II.
\vfill
\eject
\begin{center}
{\bf Table I. Characters and dimensions for some representations} \\
{\bf of the linear group $GL(N)$\\ (modified from \cite{IZ}) }
\end{center}
\vspace{0.5cm}
\noindent
\begin{tabular}{||c|c|c|c||} \hline
$\begin{array}{c}
\mbox{Young} \\
\mbox{Tableau} \\
\end{array} $
        &   $\sigma_{\{n\}}$  & $\chi_{\{n\}}(A)$ & $d_{\{n\}}$ \\ \hline
$\begin{picture}(6,6)(0,0)
\put(0,0){\framebox(6,6){$\space$}}
\end{picture}$
            &  1     &  $trA$    & $N$ \\ \hline
$\begin{picture}(12,6)(0,0)
\put(0,0){\framebox(6,6){$\space$}}
\put(6,0){\framebox(6,6){$\space$}}
\end{picture}$
            & 1    &  $\frac{1}{2} [(trA)^2 \ + \ trA^2]$  & $\frac{1}{2}
N(N+1)$ \\ \hline
$\begin{picture}(6,16)(0,0)
\put(0,0){\framebox(6,6){$\space$}}
\put(0,6){\framebox(6,6){$\space$}}
\end{picture}$
          & 1  & $\frac{1}{2} [(trA)^2 - trA^2]$  & $\frac{1}{2}
N(N-1)$ \\
\hline
$\begin{picture}(18,6)(0,0)
\put(0,0){\framebox(6,6){$\space$}}
\put(6,0){\framebox(6,6){$\space$}}
\put(12,0){\framebox(6,6){$\space$}}
\end{picture}$
           & 1 & $\frac{1}{6} [(trA)^3 + 2 trA^3 + 3 trA \ trA^2]$ &
$\frac{1}{6} N(N+1)(N+2)$ \\ \hline
$\begin{picture}(12,16)(0,0)
\put(0,0){\framebox(6,6){$\space$}}
\put(0,6){\framebox(6,6){$\space$}}
\put(6,6){\framebox(6,6){$\space$}}
\end{picture}$
           &  2 & $\frac{1}{3}[(trA)^3 - trA^3]$ & $\frac{1}{3}
N(N+1)(N-1)$ \\
\hline
$\begin{picture}(6,22)(0,0)
\put(0,0){\framebox(6,6){$\space$}}
\put(0,6){\framebox(6,6){$\space$}}
\put(0,12){\framebox(6,6){$\space$}}
\end{picture}$
    &     1 & $\frac{1}{6} [(trA)^3 + 2 trA^3 - 3 trA \ trA^2]$ &
$\frac{1}{6}
N(N-1)(N-2)$ \\ \hline
$\begin{picture}(6,6)(0,0)
\put(0,0){\framebox(6,6){$\cdot$}}
\end{picture}$
 & 1 & $trA^\dagger$  & $N$ \\ \hline
\begin{picture}(12,6)(0,0)
\put(0,0){\framebox(6,6){$\cdot$}}
\put(6,0){\framebox(6,6){$\space$}}
\end{picture}
     & - & $trA^\dagger \ trA - \frac{1}{N} trA \ A^\dagger$ &
$(N+1)(N-1)$ \\
\hline
$\begin{picture}(18,6)(0,0)
\put(0,0){\framebox(6,6){$\cdot$}}
\put(6,0){\framebox(6,6){$\space$}}
\put(12,0){\framebox(6,6){$\space$}}
\end{picture}$
     & - & $\frac{1}{2} trA^\dagger \ [(trA)^2 \ + \ trA^2 ]
-\frac{1}{N} trA \
tr(AA^\dagger)$ & $\frac{1}{2} N(N+2)(N-1)$ \\ \hline
$\begin{picture}(12,16)(0,0)
\put(0,6){\framebox(6,6){$\cdot$}}
\put(6,6){\framebox(6,6){$\space$}}
\put(6,0){\framebox(6,6){$\space$}}
\end{picture}$
     & - & $\frac{1}{2} trA^\dagger \ [(trA)^2 \ - \ trA^2 ]
-\frac{1}{N} trA \
tr(AA^\dagger)$ & $\frac{1}{2} N(N+1)(N-2)$ \\ \hline
\end{tabular}
\newpage
\begin{center}
{\bf Table II. Supercharacters for some representations of } \\
{\bf the linear supergroup  $GL(m|n)$ (constructed from \cite{IZ} and
\cite{BARS2}}) \\
\end{center}

\vspace{0.5cm}

\noindent
\begin{tabular}{||c|c|c|c||} \hline
$\begin{array}{c}
\mbox{Young} \\
\mbox{Tableau}
\end{array} $
     & $\sigma_{\{n\}}$  & $s\chi_{\{n\}}(B)$ & $sd_{\{n\}}$ \\ \hline
$\begin{picture}(6,6)(0,0)
\put(0,0){\framebox(6,6){$\space$}}
\end{picture}$
     & 1     &  $strB$    & $m+n$ \\ \hline
$\begin{picture}(12,6)(0,0)
\put(0,0){\framebox(6,6){$\space$}}
\put(6,0){\framebox(6,6){$\space$}}
\end{picture}$
     & 1    &  $\frac{1}{2} [(strB)^2 + strB^2]$  & $\frac{1}{2}
((m+n)^2+(m-n))$ \\ \hline
$\begin{picture}(6,16)(0,0)
\put(0,0){\framebox(6,6){$\space$}}
\put(0,6){\framebox(6,6){$\space$}}
\end{picture}$
     & 1     & $\frac{1}{2} [(strB)^2 - strB^2]$  & $\frac{1}{2}
((m+n)^2-(m-n))$ \\ \hline
$\begin{picture}(18,6)(0,0)
\put(0,0){\framebox(6,6){$\space$}}
\put(6,0){\framebox(6,6){$\space$}}
\put(12,0){\framebox(6,6){$\space$}}
\end{picture}$
     & 1     & $\begin{array}{l}
                \frac{1}{6} [(strB)^3 + 2 strB^3 + \\
                + 3 strB \ strB^2]
                \end{array}$ & $\frac{1}{6}
(m+n)((m+n)^2+3(m-n)+2)$ \\ \hline
$\begin{picture}(12,16)(0,0)
\put(0,0){\framebox(6,6){$\space$}}
\put(0,6){\framebox(6,6){$\space$}}
\put(6,6){\framebox(6,6){$\space$}}
\end{picture}$
     & 2    & $\frac{1}{3}[(strB)^3 - strB^3]$ & $\frac{1}{3}
(m+n)(m+n+1)(m+n-1)$ \\ \hline
$\begin{picture}(6,22)(0,0)
\put(0,0){\framebox(6,6){$\space$}}
\put(0,6){\framebox(6,6){$\space$}}
\put(0,12){\framebox(6,6){$\space$}}
\end{picture}$
     & 1  & $\begin{array}{l}
              \frac{1}{6} [(strB)^3 + 2 strB^3 - \\
              - 3 strB \ strB^2]
             \end{array}$ & $\frac{1}{6} (m+n)((m+n)^2-3(m-n)+2)$
\\ \hline
$\begin{picture}(6,6)(0,0)
\put(0,0){\framebox(6,6){$\cdot$}}
\end{picture}$
     & 1  & $strB^\dagger$  & $m+n$ \\ \hline
\begin{picture}(12,6)(0,0)
\put(0,0){\framebox(6,6){$\cdot$}}
\put(6,0){\framebox(6,6){$\space$}}
\end{picture}
     & -  & $strB^\dagger \ strB - \frac{1}{m-n} str(B \ B^\dagger)$ &
$(m+n+1)(m+n-1)$ \\ \hline
$\begin{picture}(18,6)(0,0)
\put(0,0){\framebox(6,6){$\cdot$}}
\put(6,0){\framebox(6,6){$\space$}}
\put(12,0){\framebox(6,6){$\space$}}
\end{picture}$
     & -  & $\begin{array}{l}
        \frac{1}{2} strB^\dagger [(strB)^2 + strB^2 ] - \\
       -\frac{1}{m-n} strB \ str(BB^\dagger)
        \end{array}$
                              & $\frac{1}{2}
(m+n)((m+n)^2+(m-n)-2)$ \\ \hline
$\begin{picture}(12,16)(0,0)
\put(0,6){\framebox(6,6){$\cdot$}}
\put(6,6){\framebox(6,6){$\space$}}
\put(6,0){\framebox(6,6){$\space$}}
\end{picture}$
     & -  & $\begin{array}{l}
        \frac{1}{2} strB^\dagger  [(strB)^2 - strB^2 ] - \\
        - \frac{1}{m-n} strB \ str(BB^\dagger)
        \end{array}$
                              & $\frac{1}{2}
(m+n)((m+n)^2-(m-n)-2)$ \\ \hline
\end{tabular}

\newpage

\subsection{$\alpha_{\{\dot p\}|\{q\}}$ coefficient table}
\label{Alfas}

Some examples of  the formula (\ref{alphamix}) are given in the
following table

\

\

\begin{tabular}{||c|c||} \hline
$\alpha_{\{\dot p\}|\{q\}}$
     &$ = \begin{array}{c}
          \left[ \frac{ |p|! \ |q|!}{(|p|+|q|)!} \right] ^2
          \left[ \frac{1}{\sigma_{\{p\}} \sigma_{\{q \}}} \right] ^2
          \sum_{\{t\}} \rho_{\{t\}}^{\{p\},\{q\}} \sigma_{\{t\}}^2
\alpha_{\{t\}}
\\
          \mbox{with $|t| = |p| + |q|$}
          \end{array} $ \\ \hline
$\alpha_{\begin{picture}(12,6)(0,0)
\put(0,0){\framebox(6,6){$\cdot$}}
\put(6,0){\framebox(6,6){$\space$}}
\end{picture}}$
     & $\frac{1}{4} \alpha_{\begin{picture}(12,6)(0,0)
\put(0,0){\framebox(6,6){$\space$}}
\put(6,0){\framebox(6,6){$\space$}}
\end{picture}} +  \frac{1}{4} \alpha_{\begin{picture}(6,16)(0,0)
\put(0,0){\framebox(6,6){$\space$}}
\put(0,6){\framebox(6,6){$\space$}}
\end{picture}}$    \\ \hline
$\alpha_{\begin{picture}(18,6)(0,0)
\put(0,0){\framebox(6,6){$\cdot$}}
\put(6,0){\framebox(6,6){$\space$}}
\put(12,0){\framebox(6,6){$\space$}}
\end{picture}} =
\alpha_{\begin{picture}(18,6)(0,0)
\put(0,0){\framebox(6,6){$\cdot$}}
\put(6,0){\framebox(6,6){$\cdot$}}
\put(12,0){\framebox(6,6){$\space$}}
\end{picture}} $   & $\frac{1}{9} \alpha_{\begin{picture}(18,6)(0,0)
\put(0,0){\framebox(6,6){$\space$}}
\put(6,0){\framebox(6,6){$\space$}}
\put(12,0){\framebox(6,6){$\space$}}
\end{picture}} + \frac{4}{9} \alpha_{\begin{picture}(12,16)(0,0)
\put(0,0){\framebox(6,6){$\space$}}
\put(0,6){\framebox(6,6){$\space$}}
\put(6,6){\framebox(6,6){$\space$}}
\end{picture}}$  \\ \hline
$\alpha_{\begin{picture}(12,12)(0,0)
\put(0,0){\framebox(6,6){$\cdot$}}
\put(6,0){\framebox(6,6){$\space$}}
\put(6,-6){\framebox(6,6){$\space$}}
\end{picture}} =
\alpha_{\begin{picture}(12,12)(0,0)
\put(0,0){\framebox(6,6){$\cdot$}}
\put(0,-6){\framebox(6,6){$\cdot$}}
\put(6,0){\framebox(6,6){$\space$}}
\end{picture}} $   & $\frac{4}{9} \alpha_{\begin{picture}(12,16)(0,0)
\put(0,0){\framebox(6,6){$\space$}}
\put(0,6){\framebox(6,6){$\space$}}
\put(6,6){\framebox(6,6){$\space$}}
\end{picture}} + \frac{1}{9} \alpha_{\begin{picture}(6,18)(0,0)
\put(0,0){\framebox(6,6){$\space$}}
\put(0,6){\framebox(6,6){$\space$}}
\put(0,12){\framebox(6,6){$\space$}}
\end{picture}}$  \\ \hline
$\alpha_{\begin{picture}(24,8)(0,0)
\put(0,0){\framebox(6,6){$\cdot$}}
\put(6,0){\framebox(6,6){$\cdot$}}
\put(12,0){\framebox(6,6){$\cdot$}}
\put(18,0){\framebox(6,6){$\space$}}
\end{picture}} =
\alpha_{\begin{picture}(24,8)(0,0)
\put(0,0){\framebox(6,6){$\cdot$}}
\put(6,0){\framebox(6,6){$\space$}}
\put(12,0){\framebox(6,6){$\space$}}
\put(18,0){\framebox(6,6){$\space$}}
\end{picture}}$
     &  $ \frac{1}{16} \alpha_{\begin{picture}(24,8)(0,0)
\put(0,0){\framebox(6,6){$\space$}}
\put(6,0){\framebox(6,6){$\space$}}
\put(12,0){\framebox(6,6){$\space$}}
\put(18,0){\framebox(6,6){$\space$}}
\end{picture}} + \frac{9}{16} \alpha_{\begin{picture}(18,16)(0,0)
\put(0,0){\framebox(6,6){$\space$}}
\put(0,6){\framebox(6,6){$\space$}}
\put(6,6){\framebox(6,6){$\space$}}
\put(12,6){\framebox(6,6){$\space$}}
\end{picture}}$ \\ \hline
$\alpha_{\begin{picture}(18,16)(0,0)
\put(0,6){\framebox(6,6){$\cdot$}}
\put(6,6){\framebox(6,6){$\cdot$}}
\put(6,0){\framebox(6,6){$\cdot$}}
\put(12,6){\framebox(6,6){$\space$}}
\end{picture}} =
\alpha_{\begin{picture}(18,16)(0,0)
\put(0,6){\framebox(6,6){$\cdot$}}
\put(6,6){\framebox(6,6){$\space$}}
\put(6,0){\framebox(6,6){$\space$}}
\put(12,6){\framebox(6,6){$\space$}}
\end{picture}}$
     &  $\frac{9}{64} \alpha_{\begin{picture}(18,16)(0,0)
\put(0,0){\framebox(6,6){$\space$}}
\put(0,6){\framebox(6,6){$\space$}}
\put(6,6){\framebox(6,6){$\space$}}
\put(12,6){\framebox(6,6){$\space$}}
\end{picture}} + \frac{1}{16} \alpha_{\begin{picture}(12,16)(0,0)
\put(0,0){\framebox(6,6){$\space$}}
\put(0,6){\framebox(6,6){$\space$}}
\put(6,0){\framebox(6,6){$\space$}}
\put(6,6){\framebox(6,6){$\space$}}
\end{picture}} + \frac{9}{64} \alpha_{\begin{picture}(12,22)(0,0)
\put(0,0){\framebox(6,6){$\space$}}
\put(0,6){\framebox(6,6){$\space$}}
\put(0,12){\framebox(6,6){$\space$}}
\put(6,12){\framebox(6,6){$\space$}}
\end{picture}}$  \\ \hline
$\alpha_{\begin{picture}(16,24)(0,0)
\put(0,0){\framebox(6,6){$\cdot$}}
\put(0,6){\framebox(6,6){$\cdot$}}
\put(0,12){\framebox(6,6){$\cdot$}}
\put(6,12){\framebox(6,6){$\space$}}
\end{picture}} =
\alpha_{\begin{picture}(16,24)(0,0)
\put(0,0){\framebox(6,6){$\space$}}
\put(0,6){\framebox(6,6){$\space$}}
\put(0,12){\framebox(6,6){$\space$}}
\put(-6,12){\framebox(6,6){$\cdot$}}
\end{picture}}$
     &  $ \frac{9}{16} \alpha_{\begin{picture}(16,24)(0,0)
\put(0,0){\framebox(6,6){$\space$}}
\put(0,6){\framebox(6,6){$\space$}}
\put(0,12){\framebox(6,6){$\space$}}
\put(6,12){\framebox(6,6){$\space$}}
\end{picture}} + \frac{1}{16} \alpha_{\begin{picture}(8,24)(0,0)
\put(0,0){\framebox(6,6){$\space$}}
\put(0,6){\framebox(6,6){$\space$}}
\put(0,12){\framebox(6,6){$\space$}}
\put(0,18){\framebox(6,6){$\space$}}
\end{picture}}$ \\ \hline
$\alpha_{\begin{picture}(18,16)(0,0)
\put(0,6){\framebox(6,6){$\cdot$}}
\put(6,6){\framebox(6,6){$\cdot$}}
\put(12,6){\framebox(6,6){$\space$}}
\put(12,0){\framebox(6,6){$\space$}}
\end{picture}} =
\alpha_{\begin{picture}(18,16)(0,0)
\put(0,0){\framebox(6,6){$\cdot$}}
\put(0,6){\framebox(6,6){$\cdot$}}
\put(6,6){\framebox(6,6){$\space$}}
\put(12,6){\framebox(6,6){$\space$}}
\end{picture}}$
     &  $\frac{1}{4} \alpha_{\begin{picture}(18,16)(0,0)
\put(0,0){\framebox(6,6){$\space$}}
\put(0,6){\framebox(6,6){$\space$}}
\put(6,6){\framebox(6,6){$\space$}}
\put(12,6){\framebox(6,6){$\space$}}
\end{picture}} + \frac{1}{4} \alpha_{\begin{picture}(12,22)(0,0)
\put(0,0){\framebox(6,6){$\space$}}
\put(0,6){\framebox(6,6){$\space$}}
\put(0,12){\framebox(6,6){$\space$}}
\put(6,12){\framebox(6,6){$\space$}}
\end{picture}}$  \\ \hline
$\alpha_{\begin{picture}(12,16)(0,0)
\put(0,0){\framebox(6,6){$\cdot$}}
\put(0,6){\framebox(6,6){$\cdot$}}
\put(6,0){\framebox(6,6){$\space$}}
\put(6,6){\framebox(6,6){$\space$}}
\end{picture}} $ & $\frac{1}{9} \alpha_{\begin{picture}(12,16)(0,0)
\put(0,0){\framebox(6,6){$\space$}}
\put(0,6){\framebox(6,6){$\space$}}
\put(6,0){\framebox(6,6){$\space$}}
\put(6,6){\framebox(6,6){$\space$}}
\end{picture}} + \frac{1}{4} \alpha_{\begin{picture}(12,22)(0,0)
\put(0,0){\framebox(6,6){$\space$}}
\put(0,6){\framebox(6,6){$\space$}}
\put(0,12){\framebox(6,6){$\space$}}
\put(6,12){\framebox(6,6){$\space$}}
\end{picture}} + \frac{1}{36} \alpha_{\begin{picture}(6,28)(0,0)
\put(0,0){\framebox(6,6){$\space$}}
\put(0,6){\framebox(6,6){$\space$}}
\put(0,12){\framebox(6,6){$\space$}}
\put(0,18){\framebox(6,6){$\space$}}
\end{picture}}$   \\ \hline
$\alpha_{\begin{picture}(24,10)(0,0)
\put(0,0){\framebox(6,6){$\cdot$}}
\put(6,0){\framebox(6,6){$\cdot$}}
\put(12,0){\framebox(6,6){$\space$}}
\put(18,0){\framebox(6,6){$\space$}}
\end{picture}}$
     &  $\frac{1}{9} \alpha_{\begin{picture}(12,16)(0,0)
\put(0,0){\framebox(6,6){$\space$}}
\put(0,6){\framebox(6,6){$\space$}}
\put(6,0){\framebox(6,6){$\space$}}
\put(6,6){\framebox(6,6){$\space$}}
\end{picture}} + \frac{1}{4} \alpha_{\begin{picture}(18,16)(0,0)
\put(0,0){\framebox(6,6){$\space$}}
\put(0,6){\framebox(6,6){$\space$}}
\put(6,6){\framebox(6,6){$\space$}}
\put(12,6){\framebox(6,6){$\space$}}
\end{picture}} + \frac{1}{36} \alpha_{\begin{picture}(24,6)(0,0)
\put(0,0){\framebox(6,6){$\space$}}
\put(6,0){\framebox(6,6){$\space$}}
\put(12,0){\framebox(6,6){$\space$}}
\put(18,0){\framebox(6,6){$\space$}}
\end{picture}}$   \\ \hline
\end{tabular}

\section*{Acknowledgements}

LFU acknowledges the hospitality of J.Alfaro at Universidad Cat\'olica de
Chile.
He is partially supported by the grants CONACYT 3544-E9311, CONACYT
(M\'exico)-
CONICYT (Chile) E120-2639
 and
UNAM-DGAPA-IN100694.  RM acknowledges support from a FAPESP postdoctoral
fellowship.  JA acknowledges support from the projects FON\-DE\-CYT
1950809
 and a  collaboration CONACYT(M\'exico)-CONICYT(Chile).

\end{document}